\documentclass[arxiv,Phys]{SciPost}
\pdfoutput=1 

\usepackage{mathtools}
\usepackage{environ}
\usepackage{xcolor}
\usepackage[tikz]{bclogo}
\usepackage{tikz}
\usetikzlibrary{calc}
\usepackage{amssymb,amsmath}

\def\ket#1{|{#1}\rangle}
\def\bra#1{\langle{#1}|}
\def\Tr{\text{Tr}}

\def\O{\mathbb{O}}

\newcommand{\be}[0]{\begin{equation}}
\newcommand{\ee}[0]{\end{equation}}
\newcommand{\bea}{\begin{eqnarray}}
\newcommand{\eea}{\end{eqnarray}}

\usepackage[object=vectorian]{pgfornament}

\newcommand{\sectionlineA}{%
  \vspace{.4cm}
  \noindent
  \begin{center}
  {\color{black}
    {{%
    {\begin{tikzpicture}
     \node (A) at (1,.2) {};
     \node (B) at (8,.2) {};
     \node  (C) at (0,0) {};
     \node (D) at (9,0) {};
     \draw[line width = 1pt] (A) -- (B);
     \draw[line width = 1pt] (C) -- (D);
    \end{tikzpicture}}}}
    }%
    \end{center}
  }

\newcommand{\sectionlineB}{%
  \vspace{.4cm}
  \noindent
  \begin{center}
  {\color{black}
    {{%
    {\begin{tikzpicture}
     \node (A) at (0,.2) {};
     \node (B) at (9,.2) {};
     \node  (C) at (1,0) {};
     \node (D) at (8,0) {};
     \draw[line width = 1pt] (A) -- (B);
     \draw[line width = 1pt] (C) -- (D);
    \end{tikzpicture}}}}
    }%
    \end{center}
  }

\begin{document} 
\thispagestyle{empty}

\begin{center}{\Large \textbf{
Aspects of the S transformation Bootstrap
}}\end{center}

\begin{center}
Enrico M. Brehm\textsuperscript{$\beta$},
Diptarka Das\textsuperscript{$\beta,\mu$}
\end{center}

\begin{center}
	{ ${}^{\beta}$} Max Planck Institut f\"ur Gravitationsphysik,\\
	Albert-Einstein-Institut, \\
	Potsdam-Golm, D-14476, 
	Germany.\\
	
	{${}^{\mu}$} Indian Institute of Technology, Kanpur,\\
	Department of Physics, \\
	Kanpur, 208016,
	India.\\
	
* brehm@aei.mpg.de
\end{center}

\begin{center}
\today
\end{center}


\section*{Abstract}
{\bf 
We review and systematize two (analytic) bootstrap techniques in two-dimen\-sional conformal field theories using the $S$-modular transformation. The first one gives universal results in asymptotic regimes by relating extreme temperatures. Along with the presentation of known results, we use this technique to also derive asymptotic formulae for the Zamolodchikov recursion coefficients which match previous conjectures from numerics and from Regge asymptotic analysis. The second technique focuses on intermediate temperatures. We use it to sketch a methodology to derive a bound on off-diagonal squared OPE coefficients, as well as to improve existing bounds on the spectrum in case of non-negative diagonal OPE coefficients. 
}

\vspace{10pt}
\noindent\rule{\textwidth}{1pt}
\tableofcontents\thispagestyle{fancy}
\noindent\rule{\textwidth}{1pt}
\vspace{10pt}

\section{Introduction}

\color{black}

Conformal field theories in two dimensions are among the best understood and most studied quantum field theories. 
They play a crucial role in the perturbative description of string theory and in particular appear as fixed points of renormalization group flow, describing the dynamics of statistical and condensed matter systems at criticality. 
In some cases they can even be solved exactly \cite{Belavin:1984vu} and under certain conditions -- the case of rational theories with a finite amount of primary operators -- all possible CFTs have been classified \cite{Cappelli:1986hf}. However, in the general case a full classification seems far out of reach. Nevertheless, there are certain techniques that allow to constrain the spectrum and couplings of generic irrational CFTs. Our main goal is to review two of these techniques, which are both often dubbed modular bootstrap, and thereby formulate it in such a manner that it can be used more generally. We, in particular, give many old but also new examples where these techniques lead to universal results in CFT.\footnote{Here, \textit{universal} means that the result holds in a large class of theories that is given by particular (weak) constraints to start with.} 

The techniques that we will review originally come from imposing particular modular properties on the CFTs. These are primarily the modular invariance of the torus partition function and modular covariance of torus $N$-point functions. In fact, imposing the latter for $N=1$ together with crossing symmetry on the sphere has been shown to lead to crossing symmetry and modular invariance in higher genus \cite{MOORE1988451,SONODA1988417}. 
Only then the CFT is unambiguously defined on any Riemann surface. The latter is, in particular, necessary in string theory because the perturbative description is in terms of higher genus Riemann surfaces. 
However, it is not always needed in other contexts, e.g. in condensed matter applications chiral CFTs, that often appear, are mostly not modular invariant.\footnote{A famous example of a chiral theory that is in fact modular invariant is the Monster CFT \cite{Frenkel:1988xz}.} Still, important CFTs used in low energy physics are modular invariant.

The conformal dimensions of primary operators and the coefficients of three point functions on the plane get constrained from modular properties. With these data at hand one can, in principle, construct correlation functions on any Riemann surface.\footnote{By \textit{in principle} we mean that one still needs to know the conformal blocks that are universal results to the symmetry equations (Ward identities). However, these symmetry equations are non-trivial differential equations that have not been solved in general. Or equivalently, one has to explicitly perform the sewing which is in practice not an easy task, too. } Hence, complete knowledge about these data completely determines CFT observables. An already quite established technique to (numerically) obtain a consistent set of the above data is the \textit{conformal bootstrap} that focuses mainly on the constraints coming from crossing symmetry of the four point function, which we shall not describe here, but rather refer the interested readers to \cite{Simmons-Duffin:2016gjk,Rychkov:2016iqz} and references therein. 

The setting for us, here, is a conformal field theory defined on a torus. The latter is given by its modular parameter $\tau$. All tori whose modular parameters can be transformed into each other by a modular transformation 
\begin{equation}
    \tau \mapsto \gamma\cdot \tau = \frac{a\tau+b}{c\tau+d}\,,\quad \begin{pmatrix} a & b\\ c& d\end{pmatrix}\in \mathrm{PSL}(2,\mathbb{Z})
\end{equation}
are conformally equivalent to each other.\footnote{The group $\mathrm{PSL}(2,\mathbb{Z})$ is also called the mapping class group of the torus.} A CFT should therefore \textit{look the same} on all these, which in particular means that the partition function should be invariant under these transformations. The modular group $\mathrm{PSL}(2,\mathbb{Z})$ can be generated by the two elements
\begin{align}
    S&: \tau \mapsto -\frac{1}{\tau}\,,\\
    T&: \tau \mapsto \tau+1\,.
\end{align}

\noindent 
Here, we want to focus on the first generator, $S$, and discuss how to extract or constrain data of a quantity $f$ that transforms in a particular way which will be specified in \S\ref{sec:basicSetup}. 
Following our motivation, we will choose $f$ in examples to be some quantity of the two-dimensional field theory -- a partition function, a correlation function, etc. -- and most of the time the examples that we choose shall be tied to the theory defined on a torus. 
However, neither the first nor the second is necessary to present the general techniques. This fact is aligned with the spirit of our article, namely we assume a general function $f$, that depends on some complex variables $\tau,\bar{\tau}$ and that shows some transformation property under the ``$S$-transformation'' of $\tau,\bar{\tau}$. Then if one assumes that the function can be expanded in a quite general way, using the latter transformation properties, one can obtain some knowledge about the expansion coefficients. In the examples, when we go back to explicit functions of a CFT, the expansion will be very specific and the expansion coefficients will be related to the data of the CFT. 

We will actually not work with $\tau,\bar{\tau} \in \mathbb{C}$ but shall restrict ourselves to the upper imaginary axis so that we can use the parametrization $\tau = i \frac{\beta}{L}$ and $\bar{\tau} = \tau^\star = -i\frac{\beta}{L}$, with $\beta \in \mathbb{R}^+$, furthermore we choose $L = 2\pi$. Then $f$ becomes a function of the real positive parameter $\beta$ that transforms under the $S$-transformation as 
\begin{equation}
\beta \mapsto \frac{L^2}{\beta} = \frac{4\pi^2}{\beta}\,. 
\end{equation}

\noindent 
In case of $(\tau,\bar\tau)$ being the modular parameters of a torus, the above choice picks a rectangular torus. In conformal field theory the partition function on a rectangular torus represents a critical system that lives on a circle of length $L$ in a thermal state with temperature $1/\beta$. Correlation functions on that torus are then the thermal expectation values. From now on, we generally talk of inverse temperature when talking about $\beta$ even when there is no real thermal system involved. 

The two techniques we present, work in different, quite special \textit{temperature regimes}. 
The first one is the extreme temperature regime, namely when $\beta\to 0$ and $\beta\to \infty$. 
The $S$-transformation directly relates the two regimes to each other. The high temperature behavior of $f(\beta)$ is then related to its low temperature behavior via the respective transformation property. In many cases, there exists some explicit knowledge about $f$ in one of the extreme regimes which allows to obtain information about $f$ in the other one. For a conformal theory on a torus, many quantities at low temperature are dominated by the vacuum contribution, hence are universal and under control. An approximation to the high temperature behavior of the quantity then follows by its transformation property. Already the physical intuition tells us that \textit{high energies} should contribute at high temperatures\footnote{Recently there has been active effort to make this assumption rigorous using \textit{complex Tauberian theorems}, which we briefly review in \S\ref{sec:regimes}.}. And indeed, we can extract very good approximations to high energy data of the CFT as we will see in the examples. Generally, the first technique gives \textit{approximate results on asymptotic data} that appear in the quantity $f$. 

The other special regime is the medium temperature regime. More precisely we will use that the $S$-transformation has a fixed point at $\tau = i$ or at inverse temperature $\beta = L (=2\pi)$. 
Because of this, we can construct constraint equations by applying particular functionals on $f$. 
Depending on the choice of functionals these equations are very often not handy in their bare form and can even be trivial. 
However, reformulating and combining them in particular ways, i.e. choosing the functional cleverly, can lead to interesting constraints in form of bounds on possible expansion parameters. 
In conformal theories, an analysis like this e.g. led to the result that there exists an upper bound on the gap between the vacuum and the first primary \cite{Hellerman:2009bu}. 
Generally, the second method can generate \textit{bounds on the data} that appear in the quantity $f$. 

The text is organized as follows. 
In \S\ref{sec:basicSetup} we shortly define the basic setup. 
We, in particular, specify the transformation property \eqref{eq:LinRelation} that we demand from our quantity $f$ such that the two techniques can be applied. 
Some examples in conformal field theory are given in \S\ref{sec:Examples1}. They include partition functions, grand canonical ensembles for a $\mathfrak{u}(1)_k$ current, four-point functions on the plane, four-point blocks on the plane, and correlation functions on the torus. 
In \S\ref{sec:asymptotics}, we review and systematize the first technique. An important assumption is the way the quantity $f$ can be expanded \eqref{eq:fExpansion}. The final result of this section is \eqref{eq:largnAppr} which gives an approximation to asymptotic expansion coefficients. 
The best-known application is the approximation of the density of states for large conformal dimensions \cite{Nahm:1974jm}, given in \eqref{eq:CardyFormula}. 
Other known examples are results on averaged diagonal matrix elements \cite{Kraus:2017kyl}, given in \eqref{eq:diag}, and averages over squared matrix elements from considering torus two point functions \cite{bdd}, given in \eqref{eq:off-diag}, and from the pillow four point function \cite{Das:2017vej}, given in \eqref{eq:pillow}. We here also present results on squared matrix elements in theories with a $\mathfrak{u}(1)_k$ symmetry \eqref{eq:off-diagQ}, and on Zamolodchikov recursion coefficients that appear in conformal sphere four-point blocks, \eqref{eq:cn-cont1}, \eqref{eq:cn-disc1}, and \eqref{eq:cn-disc2}. 
In \S\ref{sec:regimes} we comment on the regimes of validity and error bounds on results obtained by the above technique. 
In \S\ref{sec:midT}, we review and systematize the second technique. Again it is important to assume a similar expansion of the quantity $f$. 
Then, after constructing an invariant quantity $g$ from $f$ we basically apply the technique developed in \cite{Hellerman:2009bu} on $g$. The results of the latter paper also serve as first example in \S\ref{sec:Hellerman}:
Applying the technique on the partition function gives a universal bound on the spectrum \eqref{eq:HellermanBound}. In the second example in \S\ref{sec:T1ptfct}, we focus on the torus one-point function at large central charge with only non-negative diagonal OPE coefficients. The technique leads to an upper bound on the first gap in the spectrum given by $c/11.05$. 
In the last example in \S\ref{sec:T2ptfct}, we use the torus two-point function and, among giving weak bounds on the spectrum, present a method to obtain bounds on particular OPE coefficients. 

\section{Basic setup}\label{sec:basicSetup}

We want to consider a function $f_\alpha(\tau,\bar{\tau})$ of two complex variables $(\tau,\bar{\tau})$ and additional variables $\alpha$. The set of $\alpha$ is only restricted by the assumption that under the \textit{$S$-transformation}
\begin{equation}
\tau \mapsto - \frac{1}{\tau}\,, \quad \bar{\tau} \mapsto -\frac{1}{\bar \tau}
\end{equation}
we can define the transformation property of $f_\alpha$ as follows 
\begin{equation}\label{eq:LinRelation}
f_\alpha(\tau,\bar \tau) = \int d\gamma \, \mathbb{S}_{\alpha\gamma}(\tau,\bar\tau) f_\gamma(- \frac{1}{\tau},-\frac{1}{\bar \tau})\,,
\end{equation}
where the integral with its kernel $\mathbb{S}$ represents a linear map on the space of functions $f_\gamma$. The kernel can be regarded as a distributional object that e.g. could include $\delta$-functions in which case the integral could (partially) be replaced by a sum, too. 

\sectionlineA

\subsection{Examples} \label{sec:Examples1}

\subsubsection{Partition function}

Let us consider the partition function of a two-dimensional conformal field theory on a torus with modular parameters $(\tau,\bar{\tau})$. In this example there is no $\alpha$ dependence and we write $f(\tau,\bar \tau) = Z(\tau, \bar \tau)$. In a modular invariant theory the partition function must be invariant under the $S$-transformation, which means that \eqref{eq:LinRelation} reduces to 
\begin{equation}\label{eq:PartFct}
Z(\tau,\bar \tau) = Z\!\left(- \frac{1}{\tau},-\frac{1}{\bar \tau}\right)\,.
\end{equation}

\subsubsection{Grand canonical ensemble (GCE)} 

Let us consider the partition function on a torus with non-vanishing chemical potentials $\nu,\bar \nu$ for a $\mathfrak{u}(1)_k$ current. In this case $\alpha$ is associated with the chemical potential and we write $f_{\nu,\bar\nu}(\tau,\bar\tau) = Z(\tau,\bar\tau;\nu,\bar \nu)$.  If the ordinary partition function is supposed to be modular invariant, then the GCE transforms such that \cite{Benjamin:2016fhe,Gaberdiel:2012yb}
\begin{equation}\label{eq:GCE}
Z(\tau,\bar\tau;\nu,\bar \nu) =\exp\left[ic\pi k\left(\frac{ \bar{\nu}^2}{\bar{\tau}}-\frac{\nu^2}{\tau}\right)\right] Z\!\left(-\frac{1}{\tau},-\frac{1}{\bar\tau};\frac{\nu}{\tau},\frac{\bar\nu}{\bar\tau}\right)\,,
\end{equation}
which is of the form \eqref{eq:LinRelation} with $$\mathbb{S}_{(\nu,\bar{\nu})(\mu,\bar{\mu})}(\tau,\bar\tau) = \delta\left(\mu-\frac{\nu}{\tau}\right)\delta\left(\bar\mu-\frac{\bar\nu}{\bar\tau}\right) \exp\left[ic\pi k\left(\bar\tau \bar{\mu}^2-\tau\mu^2\right)\right].$$

\subsubsection{Four-point function on the plane} Let us consider the 4-point function on the plane 
\begin{equation}
\left\langle O(\infty) O(1) O(z) O(0)\right\rangle_\mathbb{C} 
\end{equation}
which is in particular invariant under the conformal transformation $z \mapsto 1-z$, one of the crossing symmetries of the four-point function. If we define the monome $q = e^{i\pi \tau}$ with $\tau = i \frac{K(1-z)}{K(z)}$, where $K(z)$ is the elliptic integral of first kind, then the above crossing symmetry can be regarded as invariance under the $S$-transformation of $\tau$, i.e. with $f(\tau) := \left\langle O(\infty) O(1) O(z(\tau)) O(0)\right\rangle_\mathbb{C}$ the crossing relation can be written as
\begin{equation}
    f(\tau) = f\left(-\frac{1}{\tau}\right)\,.
\end{equation}

\noindent
In the more general case of different operator insertions the above transformation is not an invariance but relates the respective correlators to each other. In that case $\alpha$ is given by the two orderings in which the operators appear in the correlator, i.e. we can define $\alpha = (ijkl)$ then
\begin{equation}
    f_{(ijkl)}(\tau) = \left\langle O_l(\infty) O_k(1) O_j(z) O_i(0)\right\rangle_\mathbb{C} 
\end{equation}
and the crossing relation can be written as
\begin{equation}
    f_{(ijkl)}(\tau) = f_{(kjil)}\left(-\frac{1}{\tau}\right)\,.
\end{equation}

\noindent
Note, that this is the first example where $\tau$ is not related to the modular parameter of a torus.

\subsubsection{Conformal four-point blocks} 

Let us consider the four point blocks $F_{12}^{34}(h;z)$ that appear in the expansion of the four point function on the complex plane. We employ the following parameterization of the CFT data: $c = 1 + 6Q^2$, $Q = b+b^{-1}$, and $h = \alpha(q-\alpha)$, where $Q$ is called ``background charge'', $b$ the ``Liouville coupling'', and $\alpha$ the ``momentum''. From crossing symmetry of the full four-point function in Liouville theory one can find invertible fusion relation \cite{ponsot1999liouville}
\begin{align}\label{eq:block-st}
F^{14}_{32}(h_t;1-z) &= \int_{\frac{Q}{2}+i \mathbb{R}} \frac{d\alpha_s}{2i} \,\mathbb{S}_{\alpha_s \alpha_t}\left[\begin{matrix}
\alpha_3 &\alpha_4\\ \alpha_1 & \alpha_2 
\end{matrix}\right] \cdot F_{12}^{34}(\alpha_s;z) \,\\
&\quad+ \sum_{\gamma_{k;m,n}<\frac{Q}{2}} \underset{\alpha_s = \gamma_{k;m,n}}{\mathrm{Res}}\left(\mathbb{S}_{\alpha_s \alpha_t}\left[\begin{matrix}
\alpha_3 &\alpha_4\\ \alpha_1 & \alpha_2 
\end{matrix}\right] \cdot F_{12}^{34}(\alpha_s;z)\right)\nonumber
\end{align} 
with some kernel $\mathbb{S}$. See Appendix \ref{app:kernel} for more on it and \ref{app:anastruc} for the definition of $\gamma_{k;m,n}$. Now, by again defining the nome $q = e^{i\pi \tau}$ with $\tau = i \frac{K(1-z)}{K(z)}$, the latter equation takes the (chiral) form of $\eqref{eq:LinRelation}$ for $\tilde{\tau} = -\frac{1}{\tau}$. \\
Although the fusion kernel had been derived some time back \cite{ponsot1999liouville, Ponsot_2001}, only recently it has found applications in analytic bootstrap \cite{Jackson_2015, Chang_2016, chang2016semiclassical, Esterlis_2016, Mertens_2017,Collier:2018exn, kusuki3}. 

\subsubsection{N-point functions on the torus} Consider the correlator of $N$ primary fields on the torus
\begin{equation}
f_{w_i}(\tau,\bar{\tau}) := \left\langle O_1(w_1)\dots O_N(w_N) \right\rangle_{\tau,\bar\tau}\,.
\end{equation}

\noindent 
The elliptic coordinates $w_i$ on the torus transform under $S$-transformation as $w_i\to \frac{w_i}{\tau}$. Using the transformation properties of the primaries and invariance of the correlator under modular transformations it follows that 
\begin{equation}\label{eq:CorrTrafo}
 \left\langle O_1(w_1)\dots O_N(w_N) \right\rangle_{\tau,\bar\tau} = \frac1{\tau^{\sum_{i=1}^N h_i} {\bar{\tau}}^{\sum_{i=1}^N \bar{h}_i}}  \left\langle O_1\left(\frac{w_1}{\tau}\right)\dots O_N\left(\frac{w_N}{\tau}\right) \right\rangle_{-\frac1\tau,-\frac1{\bar\tau}}\,
\end{equation}  
which has the form of \eqref{eq:LinRelation} with $S_{ w_i w'_i}(\tau,\bar\tau) = \prod_i \frac{\delta(w'_i-\frac{w_i}{\tau})}{\tau^{h_i}\bar{\tau}^{\bar{h}_i}}$\,.

\sectionlineB

\section{Asymptotic results from extreme temperature} \label{sec:asymptotics}

Now we want to present the first technique to obtain knowledge about asymptotic data that appear in the function $f_\alpha$. 
We assume that they are encoded in the latter function as some expansion coefficients. 
Therefore, let us assume that the function $f_\alpha$ can, in fact, be expanded schematically in $q = e^{-\beta}$ as
\begin{equation}\label{eq:fExpansion}
f_\alpha(\beta) \equiv f_\alpha(q) =   f_{\alpha,0}(q)\left(C_{\alpha,0}   + \sum_{n \ge \epsilon > 0 }\sum_{~\{m\}_n} C_{\alpha,m;n}\, f_{\alpha,m;n}(q) \right)\,
\end{equation}
with some explicit knowledge about the set of functions $f_{\alpha,m;n}(q)$ and the coefficient $C_{\alpha,0}$ but not necessarily about the other coefficients $C_{\alpha,m;n}$. 
In fact, those are the objects we want to obtain some knowledge about! The sum over values $n$ can be discrete or continuous and is distinguished from other possible expansion parameters $\{m\}_n$ by the assumption that there exist a gap $\epsilon$ to a contribution from $n=0$. 
The set $\{m\}_n$ normally depends on $n$, indicated by the subscript, and we, in particular, assume that $\{m\}_0 = \{1\}$ s.t. the $m$-sum collapses for $n=0$. An easy example is $\{m\}_n = 1$, $f_{\alpha,1;n}(q) = q^n$ and $f_{\alpha,0}(q) = e^{\beta b_\alpha}$ with some real constant $b_\alpha$. This is exactly what happens in the case of the partition function.

We further assume that in the low temperature regime $\beta \to \infty$ $(T\to 0)$ the functions behave as $f_{\alpha,n} \sim q^n $, s.t. all terms in this sum are suppressed exponentially compared to the first term. 
This allows us to write
\begin{equation}
f_\alpha(\beta) = f_{\alpha,0}(q) \left(C_{\alpha,0} + \mathcal{O}\left(e^{-\beta \epsilon}\right)\right) \stackrel{\beta\gg1}{\approx} f_{\alpha,0}(q) C_{\alpha,0} \,.
\end{equation}

\noindent 
Then, by the transformation rule \eqref{eq:LinRelation} the high temperature result $\beta \to 0$ $(T\to \infty)$ can be approximated by the low temperature result as
\begin{equation}
 f_\alpha(\beta) = \int d\gamma \, \mathbb{S}_{\alpha\gamma}(\beta) \, f_\gamma\left(\frac{4\pi^2}{\beta}\right) \stackrel{\beta\ll1}{\approx} \int d\gamma \, \mathbb{S}_{\alpha\gamma}(\beta)\, f_{\gamma,0}\left(\frac{4\pi^2}{\beta}\right) \, C_{\gamma,0}\,. \label{eq:firstAppr}
\end{equation}

\noindent
In the regime $\beta \to 0$ we demand that large $n$ contributions dominate and $\{m\}_n$ can be approximately replaced by some $\{m\}_\infty$. We then use a further approximation very familiar from statistical physics, namely replace the (possibly discrete) sums by integrals, s.t. 
\begin{equation}
 f_\alpha(\beta) \stackrel{\beta\ll1}{\approx} f_{\alpha,0}(q) \int_0^\infty dn \int dm\, C_{\alpha,m;n} \, f_{\alpha,m;n}(q) =: f_{\alpha,0}(q)\, \mathcal{I}[C_{\alpha,n}](q)\,. 
\end{equation} 

\noindent
For a convenient set of functions, $\mathcal{I}$ is some invertible integral transform of the coefficients $C_{\alpha,n}$. In the above example of $f_{\alpha,n}(q) = q^n$ it is nothing but the Laplace transform! By taking the inverse transformation and using \eqref{eq:firstAppr} we obtain 
\begin{equation}\label{eq:largnAppr}
 C_{\alpha,m;n} \approx \mathcal{I}^{-1} \left[\int d\gamma \, \mathbb{S}_{\alpha\gamma}(\beta)\, f_{\gamma,0}\left(\frac{4\pi^2}{\beta}\right) f_{\alpha,0}(\beta) \, C_{\gamma,0}\right](m;n) \,. 
\end{equation}

\noindent 
In case of $f_{\alpha,n}(q) = q^n$ the inverse Laplace transform is $\mathcal{I}^{-1}[\cdot]= \oint \frac{d\beta}{2\pi i} (\cdot) e^{\beta n}$. 

The approximation \eqref{eq:largnAppr} is the main result of this section. The quality of the approximation heavily depends on how well the sums can be approximated by integrals. In particular, it can only be good for those values of $n$ for which the main contribution to the integral transform originates from the asymptotic region $\beta\ll1$. This is typically the case for asymptotic values of $n$, too. In many cases, this can be checked by considering the saddle point approximation to the integral and demanding that the saddle for $\beta$ is small. 

\sectionlineA

\subsection{Examples}

\subsubsection{Partition function and density of states} 

The thermal partition function of a two-dimensional CFT with a non-zero gap to the unique vacuum can be expressed as
\begin{align}
Z(\beta)&= \Tr \,e^{-\beta H} = \Tr \,e^{-\beta (L_0 + \bar{L}_0 - \frac{c}{12})}\\
&= e^{\beta \frac{c}{12}} \left(1 + \sum_{\Delta\ge \epsilon >0} \rho(\Delta) e^{-\beta\Delta}\right)\,,
\end{align}
where the $\Delta$s are the conformal dimensions of non-vacuum states appearing in the spectrum, $\rho(\Delta)$ is the number of states of conformal dimension $\Delta$, $\epsilon$ is the gap in the spectrum and 1 is the contribution from the vacuum. Applying \eqref{eq:largnAppr} hence gives the high energy density of states
\begin{align}
\rho(\Delta) &\approx \frac{1}{2\pi i} \oint d\beta\, e^{\frac{\pi^2 c}{3\beta} + \beta (\Delta-\frac{c}{12})}\\
&\approx \sqrt2 \pi \left(\Delta-\frac{c}{12}\right)^{-\frac{3}{4}} e^{4\pi\sqrt{\frac{c}{12}\left(\Delta-\frac{c}{12}\right)}} \label{eq:CardyFormula}
\end{align}
also known as the Cardy density of states \cite{Cardy:1986ie, Nahm:1974jm}. 

The saddle of the integral is $\beta^\star = \frac{2\pi}{\sqrt{12\Delta/c-1}}$ which is small for $\Delta\gg \frac{c}{12}$. For generic $c>0$ and without imposing further assumptions, the Cardy density of states is a good approximation to the CFT density of states at energies $\Delta\gg \frac{c}{12}$. 
This asymptotic behaviour of the high energy density of states is of great importance for the AdS/CFT correspondence, since it matches with the horizon entropy of BTZ black holes \cite{Strominger_1998}!

\subsubsection{Thermal one-point function and averaged expectation values} 
\label{Sec:onept-tor}
The thermal expectation value of a primary operator $\mathbb{O}$ can be written as 
\begin{align}
\langle \mathbb{O} \rangle_\beta &= \Tr\, \mathbb{O} e^{-\beta H} = \sum_{i} \bra{i} \mathbb{O}\ket{i} e^{-\beta \left(\Delta_i-\frac{c}{12}\right)}\\
&\equiv  e^{\beta \frac{c}{12}}\left(\rho(\Delta_\epsilon)\overline{\bra{\Delta_\chi}\mathbb{O}\ket{\Delta_\chi}}e^{-\beta\Delta_\chi}+\sum_{\Delta\ge\Delta_\delta>\Delta_\epsilon} \rho(\Delta) \overline{\bra{\Delta}\mathbb{O}\ket{\Delta}}\, e^{-\beta \Delta}\right)\,,
\end{align}
where $\overline{\bra{\Delta}\mathbb{O}\ket{\Delta}}$ is the average over all expectation values $\bra{i}\mathbb{O}\ket{i}$ with $\Delta_i = \Delta\,$. Since the vacuum expectation value $\bra{0}\mathbb{O}\ket{0}$ vanishes, the leading contribution comes from the lightest fields in the spectrum that have non-vanishing overlap with $\mathbb{O}$. We assign the conformal dimension $\Delta_\chi$ to them. Additionally, we have to assume that there is a gap in the conformal dimensions to the next fields with conformal dimension $\Delta_\delta>\Delta_\chi$. Then, by applying \eqref{eq:largnAppr} with the transformation property \eqref{eq:CorrTrafo} we can obtain an approximation of the average expectation values (with $S = \bar{h}_\mathbb{O} -h_\mathbb{O}$) \cite{km}
\begin{align}
\overline{\bra{\Delta}\mathbb{O}\ket{\Delta}} &\approx \frac{\rho(\Delta_\chi)\overline{\bra{\Delta_\chi}\mathbb{O}\ket{\Delta_\chi}} }{\rho(\Delta)} \frac{i^S}{2\pi i} \oint d\beta \left(\frac{2\pi}{\beta}\right)^{\Delta_{\mathbb{O}}} e^{\left(\Delta-\frac{c}{12}\right)\beta - \left(\Delta_\chi-\frac{c}{12}\right)\frac{4\pi^2}{\beta}}\nonumber\\
&\approx \frac{\sqrt{2} \pi \mathcal{N}_{\O,\chi}}{\rho(\Delta)} \left(\Delta-\frac{c}{12}\right)^{\frac{\Delta_\mathbb{O}}{2}-\frac34} \exp\left(4\pi \sqrt{\left(\frac{c}{12}-\Delta_\chi\right)\left(\Delta-\frac{c}{12}\right)}\right)\,.\nonumber\\
&\approx \mathcal{N}_{\O,\chi}  \left(\Delta-\frac{c}{12}\right)^{\frac{\Delta_\mathbb{O}}{2}} \exp\left(-\frac{\pi c}{3}\left(1-\sqrt{1-\frac{12\Delta_\chi}{c}}\right)\sqrt{\frac{12\Delta}{c}-1}\right)\label{eq:diag}
\end{align}
with $\mathcal{N}_{\O,\chi} =  i^S \rho(\Delta_\chi)\overline{\bra{\Delta_\chi}\mathbb{O}\ket{\Delta_\chi}}\left(\frac{c}{12}-\Delta_\chi\right)^{\frac{\frac{1}{4}-\Delta_\mathbb{O}}{2}}$. The last step follows from using the asymptotic result for the number of states \eqref{eq:CardyFormula}. 

The saddle point of the above inverse Laplace transform is
\begin{equation}
    \beta \approx 2\pi \sqrt{\frac{\frac{c}{12}-\Delta_\chi}{\Delta-\frac{c}{12}}}+ \frac{\Delta_\O}{2\left(\Delta-\frac{c}{12}\right)}
\end{equation}
s.t. we need $\Delta\gg \frac{c}{12}$, $\Delta\gg \Delta_\O$ for the saddle point to be small and $\Delta_\chi<\frac{c}{12}$ to obtain a real one. Both requirements are necessary for \eqref{eq:diag} being a good approximation. 

The holographic description of \eqref{eq:diag} involves a Witten diagram in the BTZ geometry, of a $\phi_{\mathbb{O}}$ boundary to bulk geodesic, dual to the CFT operator $\mathbb{O}$, splitting into $\phi_\chi$ and encircling the horizon \cite{km}. The fact that this asymptotic diagonal matrix element is small, plays a key role in suppressing contributions of the primary operators in the trace distance of reduced density matrices between a pure heavy state and its corresponding thermal counterpart. Similar calculation has also been carried out in presence of a global $u(1)_k$, wherein now the OPE coefficient is dressed with Aharonov-Bohm phases \cite{Das:2017vej}.

\subsubsection{Thermal two-point function and averaged matrix elements} 
\label{Sec:twopt-tor}
The thermal two-point function of a primary operator $\O$ at two different times can be written as 
\begin{align}
\langle \mathbb{O}(t)\mathbb{O}(0) \rangle_\beta &= \Tr\left[\mathbb{O}(t)\mathbb{O}(0) e^{-\beta H}\right] \nonumber\\
&= e^{\frac{\beta c}{12}}\left(\bra{0} \mathbb{O}(t)\mathbb{O}(0) \ket{0} + \sum_{i \ge \epsilon}\bra{i} \mathbb{O}(t)\mathbb{O}(0) \ket{i} e^{-\beta \Delta}\right)\nonumber\\
&= \sum_{i,j} \left|\bra{i}\mathbb{O}\ket{j}\right|^2 e^{ i t (\Delta_i - \Delta_j)} e^{-\beta \left(\Delta_i -\frac{c}{12}\right)}\\
& \equiv \sum_{\Delta,\Delta'} \rho(\Delta)\rho(\Delta')  \overline{\left|\bra{\Delta}\mathbb{O}\ket{\Delta'}\right|^2} e^{it\omega} e^{-\beta \left(\Delta -\frac{c}{12}\right)}\,,
\end{align}
where $\omega = \Delta-\Delta'$ and $\overline{\left|\bra{\Delta}\mathbb{O}\ket{\Delta'}\right|^2}$ is the average over all squared matrix element $\left|\bra{i}\mathbb{O}\ket{j}\right|^2$ with $\Delta_i = \Delta$ and $\Delta_j = \Delta'$. The integral approximation to this double sum is a Laplace transform $\mathcal{L}$ and a Fourier transform $\mathcal{F}$. 

The leading order contribution at low temperature is given by the not normalized vacuum two-point function on a cylinder, i.e.\footnote{$(-1)^{\Delta_\mathbb{O}}$ originates from the analytic continuation to physical time $t$.}
\begin{equation}\label{eq:lowT2pt}
\langle \mathbb{O}(t)\mathbb{O}(0) \rangle_{\beta\gg 1} \approx  e^{\frac{\beta c}{12}} \bra{0} \mathbb{O}(t)\mathbb{O}(0) \ket{0} = \frac{(-1)^{-\Delta_\mathbb{O}}e^{\frac{\beta c}{12}}}{\left(2\sin\left(\frac{t}{2}\right)\right)^{2\Delta_\mathbb{O}}}\,.
\end{equation}

\noindent
Now, using \eqref{eq:largnAppr} and the transformation property \eqref{eq:CorrTrafo} we obtain \cite{bdd}
\begin{align}
\overline{\left|\bra{\Delta}\mathbb{O}\ket{\Delta + \omega}\right|^2} &= \frac{1}{\rho(\Delta)\rho(\Delta')} \mathcal{L}^{-1}_{\beta\to\Delta}\mathcal{F}^{-1}_{t\to\omega} \left[\frac{(-1)^{\Delta_\mathbb{O}} \left(\frac{\pi}{\beta}\right)^{2\Delta_\mathbb{O}} e^{\frac{\pi^2 c}{3\beta}-\frac{c\beta}{12}}}{\sinh^{2\Delta_\mathbb{O}}\!\left(\frac{\pi t}{\beta}\right)}\right]\\
&= \frac{\left(\frac{c}{12}\right)^{\frac{1}{4}} \left(\frac{12\Delta_\text{avg}}{c}-1\right)^{\Delta_\mathbb{O}-\frac{1}{2}}\rho(\Delta_\text{avg})}{\Gamma(2\Delta_\mathbb{O})\rho(\Delta)\rho(\Delta')} \left|\Gamma\!\left(\Delta_\mathbb{O} + i \frac{\omega/2}{\sqrt{12\Delta_\text{avg}/c-1}}\right)\right|^2\,,\label{eq:off-diag}
\end{align}
where $\Delta_\text{avg} = \frac{\Delta+\Delta'}{2} = \Delta+\frac{\omega}{2}$, and $\Gamma$ is the usual gamma function. 

The result \eqref{eq:off-diag} is consistent with expectations from Eigenstate Thermalization Hypothesis (see \cite{Deutsch_2018} and references therein), and additionally has a natural holographic interpretation of the absorption/decay rate of a BTZ black hole, weighted with the correct Boltzmann factors \cite{bdd,Romero-Bermudez:2018dim, Hikida:2018khg}. Further refinement of the result to just primaries have also been carried out in \cite{bdd}, which in addition to requiring $\Delta, \Delta' \gg 1$, also needs $\omega \ll \Delta, \Delta'$.

\subsubsection{GCE two-point function and averaged expectation values} 

The previous analysis can be refined when we consider the two-point function in a grand canonical ensemble of a $\mathfrak{u}(1)_k$ current (see \eqref{eq:GCE}). We denote the conserved $\mathfrak{u}(1)_k$ currents by $J_0,\bar{J}_0$. Then every state $\ket{a}$ in the spectrum carries respective quantum numbers $Q_a,\bar{Q}_a$ and we can write the two-point function in the GCE as
\begin{align}
    \left\langle O(t) O^\dagger(0)\right\rangle_{\beta,\nu,\bar\nu} &= \hspace{-.3cm}\sum_{~\Delta,\Delta',Q^+,Q^-}\hspace{-.3cm} \rho(\Delta,Q^{\pm})\rho(\Delta',Q^\pm-Q_{\mathbb{O}}^\pm) \overline{\left|\bra{\Delta,Q^\pm}\mathbb{O}\ket{\Delta',Q^\pm-Q_{\mathbb{O}}^\pm}\right|^2} \times \nonumber\\
    &\qquad\qquad \qquad \times e^{it\omega}e^{-\beta (\Delta-\frac{c}{12})}e^{2\pi i \nu_R Q^- -2\pi Q^+}\,,
\end{align}
where $\omega = \Delta-\Delta'$, $Q^\pm = Q\pm \bar{Q}$, $\nu = \nu_R + i \nu_I$\,, and $Q_{\mathbb{O}}^\pm$ are the charges of $\mathbb{O}$. 
We also use the density of states at given $\Delta$ and $Q^\pm$ that can be computed for high $\Delta$ by the present technique, too, and is given by \cite{Das:2017vej} 
\begin{equation}
    \rho(\Delta,Q^\pm) = \frac{1}{2^{3/2}}\left(\frac{c}{12}\right)^{3/4} \left(\Delta_Q -\frac{c}{12}\right)^{-5/4} e^{4\pi\sqrt{\frac{c}{12}\left(\Delta_Q-\frac{c}{12}\right)}}\,,
\end{equation}
where we use the definition $\Delta_Q := \Delta-\frac{\left(Q^+\right)^2+\left(Q^-\right)^2}{4k} = \Delta-\frac{Q^2+\bar{Q}^2}{2k}$\,. 
Finally, we define the object  $\overline{\left|\bra{\Delta,Q^\pm}\mathbb{O}\ket{\Delta',Q^\pm-Q_{\mathbb{O}}^\pm}\right|^2}$ to be the average over all squared matrix elements $\left|\bra{i}\mathbb{O}\ket{j}\right|^2$ with $\Delta_i=\Delta$, $\Delta_j = \Delta'$, $Q^\pm = Q^\pm_i$. 
Note, that the charges $Q_j^\pm$ are fixed by the fusion rules of $\mathfrak{u}(1)_k$ which essentially tells that every matrix element has to be charge neutral. 
The latter is the reason why the sum only runs over a single set of $Q^\pm$. 

The low temperature result is again given by \eqref{eq:lowT2pt} and the transformation property under $\tau \to -1/\tau =\tilde{\tau}$ follows from the transformation of the GCE \eqref{eq:GCE} and the transformation of primaries,
\begin{equation}
    \left\langle\O(\gamma w_1)\O^\dagger(\gamma w_2)\right\rangle_{\tau,\nu} = (c\tau+d)^{h} (c\bar{\tau}+d)^{\bar{h}}e^{\frac{ic\pi k \nu^2}{c\tau+d}-\frac{ic\pi k \bar{\nu}^2}{c\bar{\tau}+d}} \left\langle\O(w_1)\O^\dagger(w_2)\right\rangle_{\tilde{\tau},\tilde{\nu}}\,,
\end{equation}
which together give the high temperature approximation 
\begin{equation}
\left\langle\O(t)\O^\dagger(0)\right\rangle_{\beta\ll1} \approx e^{\frac{\pi^2 c}{3 \beta}} \frac{(-1)^{\Delta_\O}\left(\frac{\pi}{\beta}\right)^{2\Delta_\O}}{\sinh^{2\Delta_\O}\!\left(\frac{\pi t}{\beta}\right)} e^{- \frac{4\pi^2 k}{\beta} \left(\nu_R^2-\nu_I^2\right)}\,.
\end{equation}

\noindent 
Now, we can apply \eqref{eq:largnAppr} and obtain
\begin{align}
\overline{\left|\bra{\Delta,\!Q^\pm}\mathbb{O}\ket{\Delta',\!Q^\pm\!-\!Q_{\mathbb{O}}^\pm}\right|^2}        &\approx \frac{\mathcal{L}^{-1}_{\beta\to\Delta}\mathcal{F}^{-1}_{t\to\omega}\mathcal{L}^{-1}_{\nu_I\to Q^+}\mathcal{F}^{-1}_{\nu_R\to Q^-}\! \left[               e^{-\beta\frac{c}{12}}\left\langle\O(t)\O^\dagger(0)\right\rangle_{\beta\ll1}\right]}{\rho(\Delta,Q^{\pm})\rho(\Delta',Q^\pm-Q_{\mathbb{O}}^\pm)} \nonumber\\
    &\approx \frac{	2^{-\frac{3}{2}}\left(\frac{12}{c}\right)^{\Delta_\O-2} \pi}{ \left(\Delta_Q+\frac{\omega}{2}- \frac{c}{12}\right)^{\frac{1}{2}-\Delta_\O}}
	 \frac{\left|\Gamma\left(\Delta_\O +i \frac{\omega}{\sqrt{\frac{12}{c}\left(\Delta_Q+\frac{\omega}{2}-\frac{c}{12}\right)}}\right)\right|^2}{\Gamma(2\Delta_\O)}\times\nonumber\\
	&\qquad\times\frac{\rho\!\left(\Delta+\frac{\omega}{2},Q^\pm\right)}{\rho(\Delta,Q^\pm)\rho(\Delta+\omega,Q^\pm-Q^\pm_\O)}\\
    &\approx \mathcal{N} \cdot h(\Delta,Q^\pm)\cdot e^{-4\pi\sqrt{\frac{c}{12}\left(\Delta_{Q-Q_{\O}}+\frac{\omega}{2}-\frac{c}{12}\right)}}\,,
    \label{eq:off-diagQ}
\end{align}
where in the last step we took the approximation $|\omega| \ll \Delta_Q$, define $\mathcal{N} = \frac{\left(\frac{12}{c}\right)^{\Delta_\O-5/4} \pi}{ 8\,\Gamma(2\Delta_\O)}$, and 
$$
h(\Delta,Q^\pm) = \frac{\left(\Delta_Q+\frac{\omega}{2}- \frac{c}{12}\right)^{\Delta_\O-\frac{1}{2}}}{\left(\Delta_{Q-Q_\O}+\frac{\omega}{2}- \frac{c}{12}\right)^{-\frac{5}{2}}} \left|\Gamma\left(\Delta_\O +i \frac{\omega}{\sqrt{\frac{12}{c}\left(\Delta_Q+\frac{\omega}{2}-\frac{c}{12}\right)}}\right)\right|^2\,.
$$
captures the non-exponential dependence on $\Delta$ and $Q^\pm$\,.

The result \eqref{eq:off-diagQ} is the charged generalization of \eqref{eq:off-diag} and thus is relevant for ETH in presence of a global symmetry. Holographically it also probes the charged BTZ black hole, just as the diagonal result does \cite{Das:2017vej}.

\subsubsection{Four-point function on the plane and Zamolodchikov expansion} 
\label{Sec:pillow}
The $\left\langle1221 \right\rangle$ four-point function can be expanded in the Zamolodchikov blocks $H_{12}^{21}(h_i,q)$ as follows \cite{Zamolodchikov1984,Zamolodchikov1987}
\begin{align}\label{eq:1221}
f_{12}^{21}(q)&=\left\langle O_1(\infty) O_2(1) O_2(z) O_1(0)\right\rangle_\mathbb{C}  = \sum_i C_{12i}C_{i12} F_{12}^{21}(h_i,z) \bar{F}_{12}^{21}(\bar{h}_i,\bar{z})\\
&= \sum_i C_{12i}C_{i12} \Lambda_{12}^{21}(h_i,q)\bar{\Lambda}_{12}^{21}(\bar{h}_i,\bar{q})
 H_{12}^{21}(h_i,q)
 \bar{H}_{12}^{21}(\bar{h}_i,\bar{q})\,,
\end{align}
where again $q = e^{i\pi\tau} \equiv e^{-\beta/2}$ with $\tau = i\frac{K(1-z)}{K(z)}$, and we introduce
\begin{equation}
\Lambda_{kl}^{mn}(h_i,q) = (16q)^{h_i -\frac{c-1}{24}}z^{\frac{c-1}{24} - h_k -h_l} (1-z)^{\frac{c-1}{24} - h_l -h_m} \theta_3(q)^{\frac{c-1}{2}-4(h_k+h_l+h_m+h_n)}\,.
\end{equation}

\noindent
The advantage of the Zamolodchikov block is that it can be expressed in a $q$ expansion 
\begin{equation}\label{eq:Hexp}
H_{kl}^{mn}(h_i,q) = 1 +\sum_{n=1}^\infty c_n q^n,
\end{equation}
where all the dependence on the external and internal fields is encoded in the expansion coefficients $c_n = c_n(k,l,m,n,i)$. 
In addition, it can be shown that it behaves in the heavy limit, i.e. when $h\to \infty$, as \cite{kusuki2,Kusuki:2018nms}
\begin{equation}\label{eq:Hlargeh}
    H_{kl}^{mn}(h\to\infty,q)\approx  1 + \mathcal{O}(h^{-1})
\end{equation}

\noindent
The crossing relation $z\mapsto 1-z$ that we want to interpret as an $S$-transformation for $\tau$ relates the above expression to 
\begin{equation}
f_{22}^{11}(\tilde{q}) := \left\langle O_1(\infty) O_1(1) O_2(1-z) O_2(0)\right\rangle_\mathbb{C}
\end{equation}
which, due to the $\tilde{q}$ analog expansion of \eqref{eq:Hexp}, is dominated in the $\tilde{q}\to 0$, $\beta\to 0,$ limit by the vacuum contribution $h=0=\bar{h}$ and, therefore, can be approximated by
\begin{equation}
f_{22}^{11}(\tilde{q}) \approx \Lambda_{22}^{11}\!\left(0,e^{-\frac{2\pi^2}{\beta}}\right)\bar\Lambda_{22}^{11}\!\left(0,e^{-\frac{2\pi^2}{\beta}}\right) \,.
\end{equation}

\noindent
If we assume that the spectrum of primaries at high conformal dimension $\Delta=h+\bar{h}$ is dominated by fields of light conformal spin then in the $q\to 1$, $\beta \to 0,$ limit heavy channels dominate. Here we can use the $h_i^{-1}$ expansion \eqref{eq:Hexp} to approximate the Zamolodchikov recursion blocks, $H_{12}^{21}$ by 1 and \eqref{eq:1221} can be approximated by
\begin{equation}
    f_{12}^{21}(q) \approx \Lambda_{12}^{21}\!\left(0,e^{-\frac{\beta}{2}}\right)\bar\Lambda_{12}^{21}\!\left(0,e^{-\frac{\beta}{2}}\right) \int_0^\infty d\Delta\, \rho^p(\Delta) \overline{C_{12\Delta}^2} \,16^\Delta e^{-\beta \frac{\Delta}{2}}\,,
\end{equation}
where $\rho^p(\Delta) = 2\pi I_0\left(4\pi\sqrt{\frac{c-1}{12}\left(\Delta-\frac{c-1}{12}\right)}\right)$ is the density of primaries at conformal dimension $\Delta$, and $\overline{C_{12\Delta}^2}$ is the average over all OPE coefficients $C_{12\Delta_i}$ with $\Delta_i = \Delta$. This allows us to write
\begin{align}
    \int_0^\infty d\Delta\, \rho^p(\Delta) \overline{C_{12\Delta}^2} \,16^\Delta e^{-\beta \frac{\Delta}{2}} &\approx  \frac{\Lambda_{22}^{11}\!\left(0,e^{-\frac{2\pi^2}{\beta}}\right)\bar\Lambda_{22}^{11}\!\left(0,e^{-\frac{2\pi^2}{\beta}}\right)}{\Lambda_{12}^{21}\!\left(0,e^{-\frac{\beta}{2}}\right)\bar\Lambda_{12}^{21}\!\left(0,e^{-\frac{\beta}{2}}\right)}\\
    &\approx \left(\frac{\beta}{2\pi}\right)^{\frac{c-1}{2}-4(\Delta_1+\Delta_2)} e^{\frac{c-1}{12}\left(\frac{2\pi^2}{\beta} -\frac{\beta}{2}\right)} \,.
\end{align}

\noindent
Inverting the Laplace transformation on the L.H.S. then allows us to obtain \cite{Das:2017cnv, kusuki2, Collier:2018exn}
\begin{align}
    \overline{C_{12\Delta}^2} &\approx \frac{1}{16^\Delta \rho^p(\Delta)} \oint \frac{d\beta}{2\pi i}  \left(\frac{\beta}{2\pi}\right)^{\frac{c-1}2-4(\Delta_1+\Delta_2)} e^{\frac{\pi^2}{\beta} \frac{c-1}{6} +\frac{\beta}{2}(\Delta-\frac{c-1}{12})}\\
    &= \frac{1}{2\cdot16^\Delta }\left(\frac{12 \,\Delta}{c-1}-1\right)^{2(\Delta_1+\Delta_2)-\frac{c+1}{4}}\times\nonumber\\
&\qquad\times \frac{I_{4(\Delta_1+\Delta_2)-\frac{c+1}{2}}\left(2\pi\sqrt{\frac{c-1}{12}\left(\Delta-\frac{c-1}{12}\right)}\right)}{I_0\left(4\pi\sqrt{\frac{c-1}{12}\left(\Delta-\frac{c-1}{12}\right)}\right)}\\
&\approx \frac{1}{\sqrt{2}\cdot 16^\Delta} \left(\frac{12 \,\Delta}{c-1}-1\right)^{2(\Delta_1+\Delta_2)-\frac{c+1}{4}} e^{-2\pi\sqrt{\frac{c-1}{12}\left(\Delta-\frac{c-1}{12}\right)} }\,.\label{eq:pillow}
\end{align}
Holographically, the result shows that the formation and decay of black holes in a 2-2 scattering process in quantum gravity is entropically suppressed \cite{Das:2017cnv}. This suppression also sharpens the notion of OPE convergence in CFTs \cite{Fitzpatrick_2012, Pappadopulo_2012}.

The analysis can also be done with 4-point functions with arbitrary operator insertions. However, then the lightest field $\chi$ that has non-vanishing OPE coefficient with the fields fused in the $t$ channel contributes dominantly in the $\tilde{q}\to 0$, $\beta\to 0$, limit. 
The above analysis would then lead to a result for $\overline{C_{12\Delta}C_{\Delta34}}$ that depends on $C_{23\chi}C_{\chi14}$ and the conformal dimension of $\chi$.

\subsubsection{Conformal blocks and Zamolodchikov expasion} 
\label{Sec:zamu-rec}

The $s$-channel conformal block can be expressed in terms of the Zamolodchikov block as 
\begin{equation}
F_{12}^{34}(h_s,z) \equiv \Lambda_{12}^{34}(h_s,q) H_{12}^{34}(h_s,q) = \Lambda_{12}^{34}(h_s,q) \sum_{n=0}^\infty c_n^s q^n,
\end{equation}
with all the definitions as before. The $t$-channel likewise can be expressed as 
\begin{equation}
F_{32}^{14}(h_t,1-z) \equiv \Lambda_{32}^{14}(h_t,\tilde{q}) H_{32}^{14}(h_t,\tilde{q}) = \Lambda_{32}^{14}(h_t,\tilde{q}) \sum_{n=0}^\infty c_n^t \tilde{q}^n\,. 
\end{equation}

\noindent
As already mentioned in \eqref{eq:block-st} the two channels are related by
\begin{align}
F^{14}_{32}(\alpha_t;1-z) &= \int_{\frac{Q}{2}+i \mathbb{R}} \frac{d\alpha_s}{2i} \,\mathbb{S}_{\alpha_s \alpha_t}\left[\begin{matrix}
\alpha_3 &\alpha_4\\ \alpha_1 & \alpha_2 
\end{matrix}\right] \cdot F_{12}^{34}(\alpha_s;z)  \\
&\quad+ \sum_{\gamma_{k;m,n}<\frac{Q}{2}} \underset{\alpha_s = \gamma_{k;m,n}}{\mathrm{Res}}\left(\mathbb{S}_{\alpha_s \alpha_t}\left[\begin{matrix}
\alpha_3 &\alpha_4\\ \alpha_1 & \alpha_2 
\end{matrix}\right] \cdot F_{12}^{34}(\alpha_s;z)\right)
\end{align} 
via some integral kernel $\mathbb{S}.$\footnote{Remember the parametrisation $h_i = \alpha_i(Q-\alpha_i)$} Applying \eqref{eq:largnAppr} with $c_0 = 1$, and $\tilde{q} = e^{-\beta/2}$ gives for large $n$
\begin{align}
    c_n^t &\approx \oint \frac{d\beta}{2\pi i}  e^{\beta n/2} \Bigg[\int_C \frac{d\alpha_s}{2i} \,\mathbb{S}_{\alpha_s\alpha_t} \!\left[\begin{matrix} 
\alpha_3 &\alpha_4\\ \alpha_1 & \alpha_2
\end{matrix}\right] \frac{\Lambda_{12}^{34}\left(h_s,e^{-\frac{2\pi^2}{\beta}}\right)}{\Lambda_{32}^{14}\left(h_t,e^{-\frac{\beta}{2}}\right)} \\
&\hspace{2cm} + \sum_{\gamma_{k;m,n}<\frac{Q}{2}} \underset{\alpha_s = \gamma_{k;m,n}}{\mathrm{Res}}\left(\mathbb{S}_{\alpha_s \alpha_t}\left[\begin{matrix}
\alpha_3 &\alpha_4\\ \alpha_1 & \alpha_2 
\end{matrix}\right] \cdot  \frac{\Lambda_{12}^{34}\left(h_s,e^{-\frac{2\pi^2}{\beta}}\right)}{\Lambda_{32}^{14}\left(h_t,e^{-\frac{\beta}{2}}\right)}\right)\\
& \approx \oint \frac{d\beta}{2\pi i}  \int_C \frac{d\alpha_s}{2i} \,\mathbb{S}_{\alpha_s\alpha_t} \!\left[\begin{matrix} 
\alpha_3 &\alpha_4\\ \alpha_1 & \alpha_2
\end{matrix}\right] \mathfrak{h}(\beta,h_s,h_t) \nonumber\\
&\qquad+\oint \frac{d\beta}{2\pi i}  \sum_{\gamma_{k;m,n}<\frac{Q}{2}} \underset{\alpha_s = \gamma_{k;m,n}}{\mathrm{Res}}\left(\mathbb{S}_{\alpha_s \alpha_t}\left[\begin{matrix}
\alpha_3 &\alpha_4\\ \alpha_1 & \alpha_2 
\end{matrix}\right] \cdot  \mathfrak{h}(\beta,h_s,h_t)\right)
\end{align}
where we introduced $\mathfrak{h}(\beta,h_s,h_t) = 16^{h_s-h_t} \left(\frac{\beta}{2\pi}\right)^{\frac{c-1}{4}-2\sum_i h_i} e^{\frac{\beta}{2}\left(h_t+n-\frac{c-1}{24}\right)-\frac{2\pi^2}{\beta}(h_s-\frac{c-1}{24})}$ to shorten notation. 

Before we proceed further and evaluate the integrals we first want to explicitly split the $s$-channel decomposition into the contribution from the continuous spectrum $\alpha_s \in \frac{Q}{2}+i\mathbb{R}$ and a possible finite contribution from the discrete spectrum $\alpha_s \in (0,Q/2)$.\footnote{See Appendix \ref{app:kernel}.}
Let us first consider the continuous contribution. 
In that case we can write $\alpha_s = \frac{Q}{2}+i s$ and $h_s = \frac{Q^2}{4}+s^2 = \frac{c-1}{24}+s^2$. As was also shown in \cite{Collier:2018exn} the $\alpha_s$ integral is dominated by $s=0$ in the limit of $\beta \ll 1$ and given by 
\begin{align}
    c_n^{t,cont.} &\approx \left.\partial_{\alpha_s}^2\mathbb{S}_{\alpha_s\alpha_t} \right|_{\alpha_s=\frac{Q}{2}}
\oint \frac{d\beta}{2\pi i} \int_{\mathbb{R}}\frac{ds}{2} 16^{h_s-h_t} s^2 \left(\frac{\beta}{2\pi}\right)^{\frac{c-1}{4}-2\sum_i h_i} e^{\frac{\beta}{2}\left(h_t+n-\frac{c-1}{24}\right)-\frac{2\pi^2 s^2}{\beta}}\\
&\approx 16^{\frac{c-1}{24}-h_t} \frac{1}{4\pi} \left. \partial_{\alpha_s}^2\mathbb{S}_{\alpha_s\alpha_t} \right|_{\alpha_s=\frac{Q}{2}} 
\oint \frac{d\beta}{2\pi i} \left(\frac{\beta}{2\pi}\right)^{\frac{c+5}{4}-2\sum_i h_i} e^{\frac{\beta}{2}\left(h_t+n-\frac{c-1}{24}\right)}\\
&\approx \frac{16^{\frac{c-1}{24}-h_t}}2  \left. \partial_{\alpha_s}^2\mathbb{S}_{\alpha_s\alpha_t} \right|_{\alpha_s=\frac{Q}{2}} \frac{\left(2\pi\left(h_t+n-\frac{c-1}{24}\right)\right)^{2\sum_i h_i-\frac{c+9}{4}}}{\Gamma\!\left(\sum_ih_i -\frac{c+5}{24}\right)}\,.\label{eq:cn-cont1}
\end{align}

\noindent 
In particular, there is no exponential but only polynomial growth in $n$. 

As also summarized in \ref{app:anastruc}, the contribution from the discrete spectrum is in terms of a finite sum of residues at poles $\alpha_s=\alpha_1+\alpha_2 + m b < \frac{Q}{2}$, $\alpha_s=\alpha_3+\alpha_4 + mb <\frac{Q}{2}$, etc. When $\beta\ll1$, then because of $e^{-\frac{2\pi^2}{\beta}(h_s-\frac{c-1}{24})}$ the pole with lowest conformal dimension contributes most, all others are suppressed exponentially. This already tells that $m=0$. The leading contribution comes from $\alpha_s^\text{min} = \min\{\alpha_1+\alpha_2,\alpha_3+\alpha_4,\dots\}$. In case of a single pole this gives 
\begin{align}
 c_n^{t,\text{dis.}} &\approx \pi \left(\underset{\alpha_s = \alpha_s^{\text{min}}}{\mathrm{Res}} \mathbb{S}_{\alpha_s\alpha_t}\right)  16^{h_s^\text{min}-h_t} \oint \frac{d\beta}{2\pi i }\left(\frac{\beta}{2\pi}\right)^{\frac{c-1}{4}-2\sum_i h_i} e^{\frac{\beta}{2}\left(h_t+n-\frac{c-1}{24}\right)-\frac{2\pi^2}{\beta}(h_s^{\text{min}}-\frac{c-1}{24})}\\
 &\approx 2\pi^2 \left(\underset{\alpha_s = \alpha_s^{\text{min}}}{\mathrm{Res}} \mathbb{S}_{\alpha_s\alpha_t}\right)  16^{h_s^\text{min}-h_t} \left(\frac{\frac{c-1}{24}- h_s^{\text{min}}}{h_t+n-\frac{c-1}{24}}\right)^{-\frac{\nu}{2}} \times\\ 
 &\qquad \qquad\qquad\times I_\nu \left(2\pi \sqrt{\left(\frac{c-1}{24}- h_s^{\text{min}}\right)\left(h_t+n-\frac{c-1}{24}\right)}\right)\,\\
 &\approx  \mathcal{N} \cdot \left(h_t+n-\frac{c-1}{24}\right)^{\sum_i h_i -\frac{c+5}{8}} e^{2\pi \sqrt{\left(\frac{c-1}{24}- h_s^{\text{min}}\right)\left(h_t+n-\frac{c-1}{24}\right)}}\label{eq:cn-disc1}
\end{align}
where $\nu = 2\sum_i h_i -\frac{c+3}{4}$, $h_s^\text{min} = \alpha_s^\text{min}(Q-\alpha_s^\text{min})\,$, and we define $$\mathcal{N}= \pi \left(\underset{\alpha_s = \alpha_s^{\text{min}}}{\mathrm{Res}} \mathbb{S}_{\alpha_s\alpha_t}\right)  16^{h_s^\text{min}-h_t}\left(\frac{c-1}{24}- h_s^{\text{min}}\right)^{-\frac{\nu}{2}-\frac{1}{4}}\,.$$

\noindent
When $\alpha_s^\text{min}$ is not unique and when $\alpha_t \neq 0$, then there appear double poles. In that case the result depends on the coefficient of the double pole, that we call $\mathrm{dRes}$ as in \cite{Collier:2018exn}. This for example happens when $\mathbb{O}_1 = \mathbb{O}_4$ and $\mathbb{O}_2 = \mathbb{O}_3$ with $\alpha_1+\alpha_2<\frac{Q}{2}$. Then there is a double pole at $\alpha_s = \alpha_1+\alpha_2$ and the leading answer for the $\alpha_s$ integral gives
\begin{align}
    c_n^{t,\text{dis.}} &\approx \pi^3 (Q-2(\alpha_1+\alpha_2))\left(\underset{\alpha_s = \alpha_1+\alpha_2}{\mathrm{dRes}} \mathbb{S}_{\alpha_s\alpha_t}\right)  16^{h_s^\text{min}-h_t} \times\\
    &\qquad\times\oint \frac{d\beta}{2\pi i }\left(\frac{\beta}{2\pi}\right)^{\frac{c-5}{4}-2\sum_i h_i} e^{\frac{\beta}{2}\left(h_t+n-\frac{c-1}{24}\right)-\frac{2\pi^2}{\beta}(h_s^{\text{min}}-\frac{c-1}{24})}\\
    &\approx 2\pi^4 (Q-2(\alpha_1+\alpha_2))\left(\underset{\alpha_s = \alpha_1+\alpha_2}{\mathrm{dRes}} \mathbb{S}_{\alpha_s\alpha_t}\right)  16^{h_s^\text{min}-h_t} \left(\frac{\frac{c-1}{24}- h_s^{\text{min}}}{h_t+n-\frac{c-1}{24}}\right)^{-\frac{\nu'}{2}} \times\\ 
    &\qquad \qquad\qquad\times I_{\nu'} \left(2\pi \sqrt{\left(\frac{c-1}{24}- h_s^{\text{min}}\right)\left(h_t+n-\frac{c-1}{24}\right)}\right)\,\\
    &\approx  \mathcal{N}' \cdot \left(h_t+n-\frac{c-1}{24}\right)^{\sum_i h_i -\frac{c+1}{8}} e^{2\pi \sqrt{\left(\frac{c-1}{24}- h_s^{\text{min}}\right)\left(h_t+n-\frac{c-1}{24}\right)}}\,,\label{eq:cn-disc2}
\end{align}
where $\nu' = 2\sum_i h_i -\frac{c-1}{4}$, $h_s^\text{min} = (\alpha_1+\alpha_2)(Q-\alpha_1-\alpha_2) = h_1+h_2 -2\alpha_1\alpha_2\,$, and 
\begin{equation}
    \mathcal{N}' = \pi^3 \left(\underset{\alpha_s = \alpha_1+\alpha_2}{\mathrm{dRes}} \mathbb{S}_{\alpha_s\alpha_t}\right)  16^{h_s^\text{min}-h_t}\left(\frac{c-1}{24}- h_s^{\text{min}}\right)^{-\frac{\nu'}{2}-\frac{1}{4}}\,.
\end{equation}

\noindent
Note, that in case of $h_t=0$, i.e. the vacuum contribution to the $t$-channel, there is no double pole and the previous result applies. 

To conclude, the coefficients $c_n^t$ can be well approximated for large $n$. How they behave depends on the external fields. When all external fields lie in the continuous spectrum then even in the $s$-channel decomposition, only the continuous spectrum contributes. 
In that case \eqref{eq:cn-cont1} is the asymptotic result for $c_n^t$. It, in particular, does not show an exponential growth in $\sqrt{n}$. 
When one or more external fields are from the discrete spectrum, then there is a contributions from residues when any of the combinations $\{\alpha_1 + \alpha_2,\alpha_3+\alpha_4, \alpha_i \to Q - \alpha_i\}$ is smaller than $\frac{Q}{2}$. 
In that case, the field with the lowest conformal dimension contributes most and $c^t_n$ is given by \eqref{eq:cn-disc1} if $\mathbb{S}$ has a single pole.
In presence of a double pole, we give the result for $\alpha_1 = \alpha_4$ and $\alpha_2 = \alpha_3$ in \eqref{eq:cn-disc2}. Only when there is a contribution from the discrete spectrum we see an exponential growth in $\sqrt{n}$ for $c^t_n$.

We want to mention that in the case of $\alpha_i=\alpha_j \equiv \alpha$, the constrain to obtain contributions from the discrete spectrum is $2\alpha < \frac{Q}{2}$ which gives $h_\alpha = \alpha(Q-\alpha) < \frac{c-1}{32}$, and clearly the behavior of the blocks drastically changes at $h=\frac{c-1}{32}$ (this was already pointed out in \cite{kusuki3,Collier:2018exn}). 

In all cases, explicit dependence on $\mathbb{S}$ appears as a prefactor which looks in general quite complicated. In particular for $\alpha_1 = \alpha_4$ and $\alpha_2 = \alpha_3$, many of these can be found explicitly in \cite{Collier:2018exn}. We would like to emphasize that while, the monodromy matrix approach using Regge asymptotics \cite{kusuki3} reproduces the same behaviors, our approach is more direct, especially in accessing subleading contributions, primarily because it is amenable to a complex Tauberian analysis, which we review briefly next.

\sectionlineB

\subsection{Regime of validity and bounds on errors}\label{sec:regimes}
It turns out that the techniques of S-modular bootstrap in extreme temperatures, contain secretly quite a few assumptions, which if not understood rigorously can lead us to uncontrolled errors and prevent us from appreciating the uniqueness of the universal asymptotic results including the conditions under which they are valid.
\begin{enumerate}
\item As a consequence of modular invariance, \cite{hks} was able to show that, provided the spectrum is sparse enough, the high temperature partition function is dominated by the S-transformed vacuum contribution. The analysis quantifies the meaning of this sparseness. A similar analysis has also been carried out for the torus correlation functions in \cite{Kraus:2017kyl}, and conditions for vacuum dominance can also be quantified in those examples. For instance, in the low temperature regime ($\beta > 2\pi$), the above analysis gives a bound of the form,
\begin{align}\label{eq:GH}
G_H(\beta) \leq \frac{\delta}{1-\delta}G_L(\beta).
\end{align}
In the above, $G_H (G_L)$ denotes the contribution of the heavy(light) sector to the physical quantity under consideration. The quantity $\delta$ is exponentially suppressed in $\Delta_H - \tfrac{c}{12}$, where $\Delta_H$ denotes the border between the light and the heavy sectors. Next, at large $c$, from the light sector one can further isolate the vacuum contribution by requiring a sparseness condition on the light sector, which is quite relaxed, $\rho_L(\Delta) \lesssim e^{2\pi \Delta}$, for $\Delta<\Delta_H$. 
\item Most of the S-transformation applications (e.g., bootstrapping asymptotic density of states, asymptotic OPE coefficients) need the assumption that at high temperatures, high energy states give the dominant contribution. This is then used with the modular transformation to get the asymptotic contribution from the vacuum /light spectra. The assumption is actually tied very closely with the approximation of a discrete distribution\footnote{ e.g., the density of states is a sum of Dirac delta functions. } by a smooth function. 
\item What lies at the heart of the Cardy like asymptotic formulae is the assumption that the errors in the inverse Laplace transformation is bounded as well, which results in approximating a discrete distribution by a continuous one. Tauberian theorems help in estimating these errors. In particular, provided that vacuum dominance in the sense of (1) holds, the errors can be shown to be bounded in comparison with the leading asymptotic result where contribution(s) come from the light sector of the theory.
\item One starts by bounding the integrated density centered at some heavy $\Delta \rightarrow \infty$ by convolution of the physical quantity $G(\beta)$ from \eqref{eq:GH} and bandlimited functions $\phi_\pm$, 
\begin{align}
    G(\beta+it)\circledast \phi_-(t) \leq \text{Integrated density} \leq G(\beta+it)\circledast \phi_+(t).
\end{align}
Next one S-transforms the quantity, $G(\beta + it) \rightarrow G(\frac{4\pi^2}{\beta + it } )$ and separates it out into $G_L$ and $G_H$. For the light sector part, one does the integrals of the convolution by saddle point. The optimum saddle, which sets $\beta \propto \frac{1}{\Delta^\theta}$, with $\theta >0$, shows that in the large $\Delta$ regime, one has small $\beta$, which implies one has a large argument for the function $G$. This light part gives the standard Cardy like formulae for the asymptotic density of states, which gets contribution from vacuum and/or the light sectors. For the heavy part, which is still left to be dealt with, one can use \eqref{eq:GH} to show that this is suppressed provided the conditions of (1.) are satisfied. 
\end{enumerate}

\noindent
When $G(\beta)$ is the partition function this analysis was carried out by \cite{mz}. As a consequence of their analysis one can also obtain upper bound on the spectral spacings. The spacing between Virasoro primaries was upper bounded by $\approx 1.1$ in \cite{mz}, which was improved to $\approx 1$ in \cite{pg}. Tauberian analysis was further successfully applied to the case of $G(\beta) = \langle O \rangle_\beta$ in \cite{sridip}, wherein errors to \eqref{eq:diag} have been bounded. Tauberian analysis can also be carried out for the asymptotic formulae obtained for OPE coefficients starting from the pillow geometry \eqref{eq:pillow} as well as $\langle O O \rangle_\beta$ \eqref{eq:off-diag} \cite{progress}. It will be interesting to also extend these analyses to the charged partition function, however one will first need to develop an analog of \eqref{eq:GH} which will now possess additional subtleties since one needs to separate out large and low charge sectors as well. The analysis in the case of the conformal blocks,  \eqref{eq:pillow}, \eqref{eq:cn-disc1}, \eqref{eq:cn-cont1}, \eqref{eq:cn-disc2} should also give upper bounds on the weighted spectral spacings which could be interesting holographically.

\section{Constraints from intermediate temperature}\label{sec:midT}

Now we want to present the second technique to extract data about $f$. The general idea was first presented in \cite{Hellerman:2009bu}. 
Invariance of the partition function $Z(\tau)$ under modular $S$-transformation was used to obtain an upper bound on the gap between the vacuum and the first excited state. 
The main feature that was exploited is that $S$ has a fixed point at $\tau^{f.p.} = i$. Hence, because of the invariance of the partition function, $Z(\tau^{f.p})$ is a fixed point under $S$, too. 

So first, from $f_\alpha$ we want to construct a function $g$ that is invariant under the $S$-transformation, too. To do so we write 
\begin{equation}
g(\tau,\bar\tau) := \int d\gamma \, \mathbb{J}_{\gamma}(\tau,\bar\tau)\, f_\gamma(\tau,\bar\tau)\,,
\end{equation}
with some (unknown) integral kernel $\mathbb{J}_{\gamma}(\tau,\bar\tau)$. The requirement of invariance of $g$ then tells us that 
\begin{align}
g(\tau,\bar\tau) :=& \int d\gamma \, \mathbb{J}_{\gamma}(\tau,\bar\tau) \,f_\gamma(\tau,\bar\tau) \\
 =& \int d\eta\, d\gamma \, \mathbb{J}_{\gamma}(\tau,\bar\tau) \,\mathbb{S}_{\gamma\eta}(\tau,\bar\tau) \,f_\eta\!\left(-\frac1\tau,-\frac1{\bar\tau}\right)\\
 \stackrel{!}{=} g\left(-\frac1\tau,-\frac1{\bar\tau}\right) = & \int d\eta \, \mathbb{J}_{\eta}\left(-\frac1\tau,-\frac1{\bar\tau}\right) \,f_\eta\left(-\frac1\tau,-\frac1{\bar\tau}\right)\,,
\end{align}
which is satisfied if 
\begin{equation}\label{eq:Jconstraint}
\int d\gamma \, \mathbb{J}_{\gamma}(\tau,\bar\tau)\, \mathbb{S}_{\gamma\eta}(\tau,\bar\tau) = \mathbb{J}_{\eta}\left(-\frac1\tau,-\frac1{\bar\tau}\right)\,.
\end{equation}

\noindent 
Any integration kernel that shows this transformations property, which is obviously very similar to the property of $f$ itself, does the job and can be used to construct a modular invariant quantity out of $f_\alpha$. 

\sectionlineA

\subsection{Examples}

\subsubsection{N-point functions on the torus} 

In case of the $N$-point function on a torus, the behavior under $S$ transformation is specified by $\mathbb{S}_{w_iw'_i}(\tau,\bar\tau)  =\prod_i \frac{\delta(w'_i-\frac{w_i}{\tau})}{\tau^{h_i}\bar\tau^{\bar{h}_i}}$. One can simply choose $\mathbb{J}$ to have trivial dependence on the coordinates. We can further choose $w_1 =0$ due to the shift symmetry on the torus. Since $w_1 = 0$ is a fixed point under the $S$-transformation already we do not need to integrate over it. Then \eqref{eq:Jconstraint} simplifies to 
\begin{equation}
\frac{\mathbb{J}(\tau,\bar\tau)}{\tau^{h_1}\bar{\tau}^{\bar{h}_1}} \int_{T^2_\tau}\prod_{i>1}  d^2w_i\frac{ \delta(w'_i-\frac{w_i}{\tau})}{\tau^{h_i}\bar\tau^{\bar{h}_i}}  = \frac{\mathbb{J}(\tau,\bar\tau)}{\tau\bar{\tau}} \prod\limits_i\frac{1}{ \tau^{h_i-1}\bar\tau^{\bar{h}_i-1}}=\mathbb{J}\left(-\frac1\tau,-\frac1{\bar\tau}\right)\,,
\end{equation}
where we used the delta-function identity $\delta(ax) = \frac{1}{|a|} \delta(x)$.

Using the Dedekind $\eta$-function, that transforms as
\begin{equation}
\eta(\tau) = \sqrt{-i\tau} \,\eta\!\left(-\frac1\tau\right)\,,\quad \eta(\bar\tau) = \sqrt{i\bar\tau}\, \eta\!\left(-\frac1{\bar\tau}\right)\,,
\end{equation}
we can choose
\begin{equation}
\mathbb{J}(\tau,\bar\tau) = \eta(\tau)^2\eta(\bar{\tau})^2 \prod_i \eta(\tau)^{2(h_i-1)}\eta(\bar{\tau})^{2\bar (h_i-1)}
\end{equation}
to solve this equation. The $S$ invariant function expressed in terms of the inverse temperature $\beta$ is then
\begin{equation}
g(\beta) = \eta(\beta)^4\prod_i \eta(\beta)^{2(h_i+\bar h_i-2)}  \int \prod_{i\ge2} d^2w_i \left\langle \mathbb{O}_1(0,0)\mathbb{O}_2(w_2,\bar w_2)\dots \mathbb{O}_N(w_N,\bar w_N) \right\rangle_\beta\,.
\end{equation}

\noindent 
We want to point out that the above integral in the presented form generically suffers from UV divergences. A regularization scheme has to be chosen such that it does not spoil the modular properties.\footnote{We want to thank Sridip Pal for rasing this issue.} 

In case of scalar operators $\mathbb{O}_i$, i.e. when $S_i = \bar h_i -h_i = 0$, we can even choose the simpler function $\mathbb{J}(\beta) = \frac{\beta}{2\pi}\prod_i \left(\frac{\beta}{2\pi}\right)^{h_i-1}$ which gives the $S$-invariant function 
\begin{equation}\label{eq:modInvNpointEasy}
g(\beta) = \frac{\beta}{2\pi}\prod_i \left(\frac{\beta}{2\pi}\right)^{h_i-1}  \int \prod_{i\ge2} d^2w_i \left\langle \mathbb{O}_1(0,0)\mathbb{O}_2(w_2,\bar w_2)\dots \mathbb{O}_N(w_N,\bar w_N) \right\rangle_\beta\,.
\end{equation}

\subsubsection{Grand canonical ensemble (GCE)} 

In case of the GCE the $S$-transformation kernel is given by 
$$\mathbb{S}_{(\nu,\bar{\nu})(\mu,\bar{\mu})}(\tau,\bar\tau) = \delta\left(\mu-\frac{\nu}{\tau}\right)\delta\left(\bar\mu-\frac{\bar\nu}{\bar\tau}\right) \exp\left[ic\pi k\left(\bar\tau \bar{\mu}^2-\tau\mu^2\right)\right]$$ 
such that $\mathbb J$ should satisfy
\begin{equation}\label{eq:GGEinv}
\tau\bar{\tau}\exp\left[ic\pi k\left(\bar\tau\bar{\mu}-{\tau}{\mu}^2\right)\right]\, \mathbb{J}_{\left({\tau}{\mu},{{\bar{\tau}\bar{\mu}}}\right)}(\tau,\bar\tau) = \mathbb{J}_{(\mu,\bar{\mu})}\left(-\frac1\tau,-\frac1{\bar\tau}\right)\,.
\end{equation}

\noindent 
To construct such a function we can consider the generalized theta function (with $q = e^{2\pi i\tau}$)
\begin{align}
\theta[a,b](\tau,z) = \sum_{n\in\mathbb{Z}} q^{\frac{1}{2}(n+a)^2}e^{2\pi i (n+a)(z+b)}
\end{align}
which transforms under $S$ transformationa as
\begin{equation}
\theta[a,b]\left(-\frac1\tau,\frac{z}{\tau}\right) = \sqrt{-i\tau} e^{2\pi i a b +i\pi\frac{z^2}{\tau}}\theta[b,-a](\tau,z)\,.
\end{equation}

\noindent
Taking the complex conjugate gives
\begin{equation}
\theta[a,b]\left(-\frac1{\bar\tau},\frac{\bar z}{\bar\tau}\right) = \sqrt{i\bar\tau} e^{-2\pi i a b -i\pi\frac{\bar z^2}{\bar\tau}}\theta[b,-a](\bar\tau,\bar z)\,.
\end{equation}

\noindent 
Hence, if we define 
\begin{equation}
J_{\nu,\bar\nu}(\tau,\bar{\tau}) = \eta(\tau)^2 \eta(\bar\tau)^2\left[\frac{\theta[0,0]\!\left(\tau,\nu\right) \theta[0,0]\!\left({\bar\tau},\bar \nu\right)}{\eta\left({\tau}\right)\eta\left({\bar\tau}\right)}\right]^{ck}
\end{equation}
it fulfills \eqref{eq:GGEinv}.

\sectionlineB

\noindent 
Having an $S$ invariant function $g(\beta)$ at hand, we can follow the logic of \cite{Hellerman:2009bu} to obtain some constraints on the expansion coefficients of $f$. Therefore we assume an expansion similar but a bit simpler than the one used in \eqref{eq:fExpansion},
\begin{equation}
f_\alpha(\beta) \equiv f_\alpha(q) =  \sum_{n \ge 0 } C_{\alpha,n}\, f_{\alpha,n}(q) \,,
\end{equation}
such that we can write
\begin{equation}\label{eq:gExpansion}
g(\beta) = \sum_{n\ge 0} \int d\gamma \, C_{\gamma,n}\, \mathbb{J}_{\gamma}(\beta) \,f_{\gamma,n}(\beta)
\end{equation}

\noindent
It is clear that the $S$ invariant function $g$ has a fixed point at $\beta^\star = 2\pi$ ($\tau^{f.p.} = i$). Now consider any operator $\mathcal{D}_\beta$ that is odd under the $S$ transformation, i.e. it satisfies
\begin{equation}
\mathcal{D}_\beta = - \mathcal{D}_{\frac{4\pi^2}{\beta}}\,,
\end{equation} 
and consider the corresponding functional 
\begin{equation}
F[\cdot] := \lim_{\beta\to \beta^\star} \mathcal{D}_\beta (\cdot)\,.
\end{equation}

\noindent
By construction, $F$ maps any $S$-invariant function to zero, s.t. we obtain
\begin{equation}\label{eq:fctalConstraint}
F[g] \equiv 0\,.
\end{equation}

\noindent 
Using \eqref{eq:fExpansion} we can write this as 
\begin{equation}\label{eq:constraints2}
0 = \sum_{n\ge 0} \int d\gamma \, C_{\gamma,n}\, F[\mathbb{J}_{\gamma}(\beta) \,f_{\gamma,n}(\beta)]
\end{equation}

\noindent
Simple examples for odd operators are the derivatives $D_\beta^p := \left(\beta \frac{\partial}{\partial\beta}\right)^p$ with odd $p$, and any linear combination of these.  

Now, the general idea is to interpret \eqref{eq:fctalConstraint} as constraint equations on $g$ and via \eqref{eq:constraints2} on the expansion coefficients of $f_\alpha$. There are infinitely many of these constraints because one can construct infinitely many linear independent odd operators. The goal is to reformulate these constraints in such a way that one obtains information on particular data of the function. For example, if the index set of $n$ is not known, one can construct functionals such that 
\begin{align*}
\int d\gamma \, C_{\gamma,n}\, F[\mathbb{J}_{\gamma}(\beta) \,f_{\gamma,n}(\beta)] < 0 
\end{align*}
only in a finite region of all possible $n$. Then it is necessary that there must be some contribution to $f$ from that region. The functional $F$ that minimizes this region is then optimal in that sense. 

How to choose a functional $F$ and how to obtain some bounds from \eqref{eq:constraints2} depends on the knowledge about $f$ and very often depends on additional assumptions. Here,  examples are the most illuminating.  

\sectionlineA

\subsection{Examples}

\subsubsection{Partition function and bounds on the spectrum}\label{sec:Hellerman}

This example originates from \cite{Hellerman:2009bu, Hellerman:2010qd} and gives a universal constraint on the spectrum of a modular invariant conformal field theory. By assumption the partition function is modular invariant, so there is no need to construct an invariant function. The respective thermal partition function can be expressed as
\begin{align}
Z(\beta)&= \Tr \,e^{-\beta H} = \Tr \,e^{-\beta (L_0 + \bar{L}_0 - \frac{c}{12})}\\
&=  e^{\beta \frac{c}{12}} + \sum_{\Delta\ge \epsilon >0} \rho(\Delta) e^{-\beta\left(\Delta-\frac{c}{12}\right)}\,.
\end{align}

\noindent 
Let us choose as functionals the above examples $F_p[\cdot] = \lim\limits_{\beta\to\beta^\star} D_\beta^p$ which gives the constraint equations
\begin{equation}
0 =B_p\!\left(\frac{\pi c}{6}\right)e^{ \frac{\pi c}{6}} + \sum_{\Delta\ge \epsilon >0} \rho(\Delta)\, B_p\!\left(-2\pi\Delta+\frac{\pi c}{6}\right) e^{-2\pi\left(\Delta-\frac{c}{12}\right)}\,,
\end{equation}
where $B_p(x)$ are the Bell polynomials which can be expressed as
\begin{equation}
B_p(x) = \sum_{n=1}^p S_p^{(n)} x^n
\end{equation}
with $S_p^{(n)}$ being the Stirling number of second kind that gives the number of ways of partitioning a set of $p$ elements into $n$ non-empty subsets. For example
\begin{align}
B_1(x) &= x\,,\\
B_3(x) &= x^3 +3x^2 +x\,.
\end{align}

\noindent
These two already suffice to obtain some knowledge about the spectrum of the theory, too. Therefore, consider the functional $F_3[Z]+a F_1[Z]$ which gives
\begin{align}\label{eq:constraint3}
0 &= \left(B_3\!\left(\frac{\pi c}{6}\right)+a B_1\!\left(\frac{\pi c}{6}\right)\right)e^{ \frac{\pi c}{6}} +\\
& \quad+ \sum_{\Delta\ge \epsilon >0} \rho(\Delta) \left(B_3\!\left(-2\pi\Delta+\frac{\pi c}{6}\right)+a B_1\!\left(-2\pi\Delta+\frac{\pi c}{6}\right)\right) e^{-2\pi\left(\Delta-\frac{c}{12}\right)}\nonumber
\end{align}
for all values of $a$. The polynomial $B_3\!\left(-2\pi\Delta+\frac{\pi c}{6}\right)+a B_1\!\left(-2\pi\Delta+\frac{\pi c}{6}\right)$ goes to $-\infty$ for large $\Delta$ and has three zeros at
\begin{equation}
\Delta_0 = \frac{c}{12}\,,\quad \Delta_\pm = \frac{c}{12}+\frac{9\pm 3\sqrt{5-4a}}{12\pi}\,.
\end{equation}

\noindent 
Now, one can choose ''$a$`` such that the vacuum, i.e. $\Delta= 0$, gives a non-positive contribution which is the case for $a\le -\frac{B_3(\tfrac{\pi c}{6} )}{B_1(\tfrac{\pi c}{6})} = -1 - \frac{c\pi(18+c\pi)}{36}$. Then, only $\Delta_0<\Delta<\Delta_+$ can give positive contribution to the sum on the r.h.s of \eqref{eq:constraint3}. There must be some contribution from that region to cancel the negative contribution from any other possible $\Delta$. The region is minimal for $a = -\frac{B_3\!\left(\frac{\pi c}{6}\right)}{B_1\!\left(\frac{\pi c}{6}\right)}$ which gives $\Delta_+ = \frac{c}{6} + \frac{3}{2\pi}$. Hence, a consistent spectrum of a modular invariant CFT has to have at least one field in the range
\begin{equation}\label{eq:HellermanBound}
\frac{c}{12} < \Delta < \frac{c}{6} + \frac{3}{2\pi}\,.
\end{equation}

\noindent
In particular, if $c< 12-\frac{9}{\pi}$ then $\Delta_+< 2$ which excludes that the field in the above range is a descendant of the identity. This means there either is a primary in the above region or there is a primary with conformal wight $\Delta < \frac{c}{6} + \frac{3}{2\pi}-1$ such that its first descendants are in that region. It, in particular, gives an upper bound on a possible gap in the theory! 

This first example was already refined to the case of the spectrum of primaries in \cite{Hellerman:2009bu}. It was shown that one gets an (improved) upper bound on the gap to the first primary in a generic unitary modular invariant CFT by expanding the partition function in terms of characters. It is given by $\Delta < \frac{c}{6} + 0.473695$. Ever since, the bound has been improved by optimizing the above functional \cite{Friedan:2013cba,Collier:2016cls}. The respective numerical method to find the best functional is called linear programming \cite{Rattazzi:2008pe}. 

Under additional assumption one can go further and, e.g. bound state degeneracies at particular (low) conformal dimensions. This has been done in \cite{Hellerman:2010qd} where they construct an upper bound on the number of marginal deformations under the assumption of cluster decomposition, the absence of a continuum of states just above the vacuum, or even better with no relevant operator, and $c< 24$. This can for example bound dimensions of moduli spaces of Calabi-Yau manifolds (see also \cite{Fiset:2015pta}). 

The present technique and applications thereof have also been used in \cite{Afkhami-Jeddi:2019zci,Hartman:2019pcd,Benjamin:2019stq,Keller:2012mr,Qualls:2014oea,Cho:2017fzo,Anous:2018hjh,Ashrafi:2016mns,Ashrafi:2019ebi}. 

\subsubsection{Constraints from torus one-point functions at large central charge} \label{sec:T1ptfct}

As mentioned earlier, sometimes it is convenient to impose further assumptions to obtain a manageable set of consistency conditions. For the present example, we want to impose two assumptions. 
Firstly, we take the central charge $c$ to be large and, secondly, we assume that diagonal OPE coefficients $C^{\mathbb{O}}_{\mathbb{O}_i\mathbb{O}_i}$ are non-negative. The latter can e.g. be analysed in Liouville theory. There the OPE coefficients are given by the DOZZ formula \cite{Dorn:1994xn,Zamolodchikov:1995aa}. In the continuous (physical) spectrum, i.e. when $h\ge \frac{c-1}{24}$, $C^{\mathbb{O}}_{\mathbb{O}_i\mathbb{O}_i}$ has neither zeroes nor poles as a function of $\Delta_i$. Hence, it does not change sign and we can choose $\O$ s.t. it is always positive.\footnote{We want to thank Sylvain Ribault for giving us insight into this.} However, we cannot assure that OPE coefficients behave the same way in generic CFTs. 

Now, let us consider the torus one-point function
\begin{equation}
\left\langle \mathbb{O}\right\rangle_\tau = \sum_{\mathbb{O}_i} C^{\mathbb{O}}_{\mathbb{O}_i\mathbb{O}_i} F^{h_i}_{c,h_\mathbb{O}}(q) F^{\bar h_i}_{c,\bar{h}_\mathbb{O}}(\bar q) 
\end{equation}
expanded in the one-point conformal torus blocks $F^{h_i}_{c,h_\mathbb{O}}$. For generic central charge $c$, the coefficients of their $q$ ($\bar{q}$) expansion can be computed from a recursive formula \cite{Hadasz:2009db}. However, in large $c$ they are known explicitly and are given by
\begin{equation}
 F^{h_i}_{c,h_\mathbb{O}}(q) \stackrel{c\gg1}{=} \frac{q^{h_i-\frac{c}{24}}}{1-q}  \,_2F_1\!\left(h_\mathbb{O},1-h_\mathbb{O};2h_i;\frac{q}{q-1}\right)\,.
\end{equation}

\noindent
Therefore, up to large $c$ corrections and with $h_\mathbb{O} =\bar{h}_\mathbb{O}$, we can define a modular invariant function 
\begin{align}
g(\beta) 	&= \left(\frac{\beta}{2\pi}\right)^{h_\mathbb{O}} \left\langle \mathbb{O}\right\rangle_\tau\\
			&= \sum_i  C^{\mathbb{O}}_{\mathbb{O}_i\mathbb{O}_i}  \frac{e^{-\beta\left(h_i+\bar{h}_i -\frac{c}{12}\right)} \left(\frac{\beta}{2\pi}\right)^{h_\mathbb{O}}}{(1-e^{-\beta})^2}  \times \label{eq:1PtTorusBlock}\\
			&\qquad\quad\times\,_2F_1\!\left(h_\mathbb{O},1\!-\!h_\mathbb{O};2h_i;\frac{e^{-\beta}}{e^{-\beta}-1}\right) \,_2F_1\!\left(h_\mathbb{O},1\!-\!h_\mathbb{O};2\bar{h}_i;\frac{e^{-\beta}}{e^{-\beta}-1}\right)\,.\nonumber
\end{align}

\noindent
Surprisingly, considering only the simple functional $F_1[\cdot] = \lim_{\beta\to2\pi} \beta \partial_\beta (\cdot)$ already gives a quite strong constraint. So applying $F_1$ on $g$ gives
\begin{equation}
F_1[g] = \sum_i  C^{\mathbb{O}}_{\mathbb{O}_i\mathbb{O}_i} \mathfrak{f}_{F_1}(h_\mathbb{O},h_i,\bar{h}_i,c)\,,
\end{equation}
where $\mathfrak{f}_{F_1}$ is given in Appendix \ref{app:OnePtConstraint}. 
For fixed $h_\mathbb{O}$ and $c\gg1$, $\mathfrak{f}_{F_1}$ can be interpreted as a function from $\mathbb{R}^2_+$ to $\mathbb{R}$, where $\mathbb{R}^2_+$ is the space of all possible conformal weights $(h,\bar{h})$. The set of points that are mapped to zero, i.e. $\mathfrak{f}^{-1}_{F_1}(0)$ -- also called zero variety, separates the region of possible positive and negative contributions to the above sum. We were not able to handle $\mathfrak{f}^{-1}_{F_1}(0)$ analytically. However, the numerical study of the problem is feasible. It turns out that
\begin{itemize}
\item $\mathfrak{f}^{-1}_{F_1}(0)$ is connected in $\mathbb{R}^2_+$ and intersects the two axes. This means that there is a finite region in $\mathbb{R}^2_+$ bounded by $\mathfrak{f}^{-1}_{F_1}(0)$ and the two axes. The contribution from that region has to compensate the contribution from its (non-compact) complementary region. 
\item The region grows monotonically with larger ${h}_\mathbb{O}$. For small $h_\mathbb{O}$, however, $\mathfrak{f}^{-1}_{F_1}(0)$ never goes below $h_i+\bar{h}_i = \frac{c}{12}$ and only for very large $h_\mathbb{O}$ grows above $h_i+\bar{h}_i = \frac{c}{6}$, i.e. the Hellerman bound \cite{Hellerman:2009bu}. 
\end{itemize}

\begin{figure}[ht]
\centering
\includegraphics[scale=1]{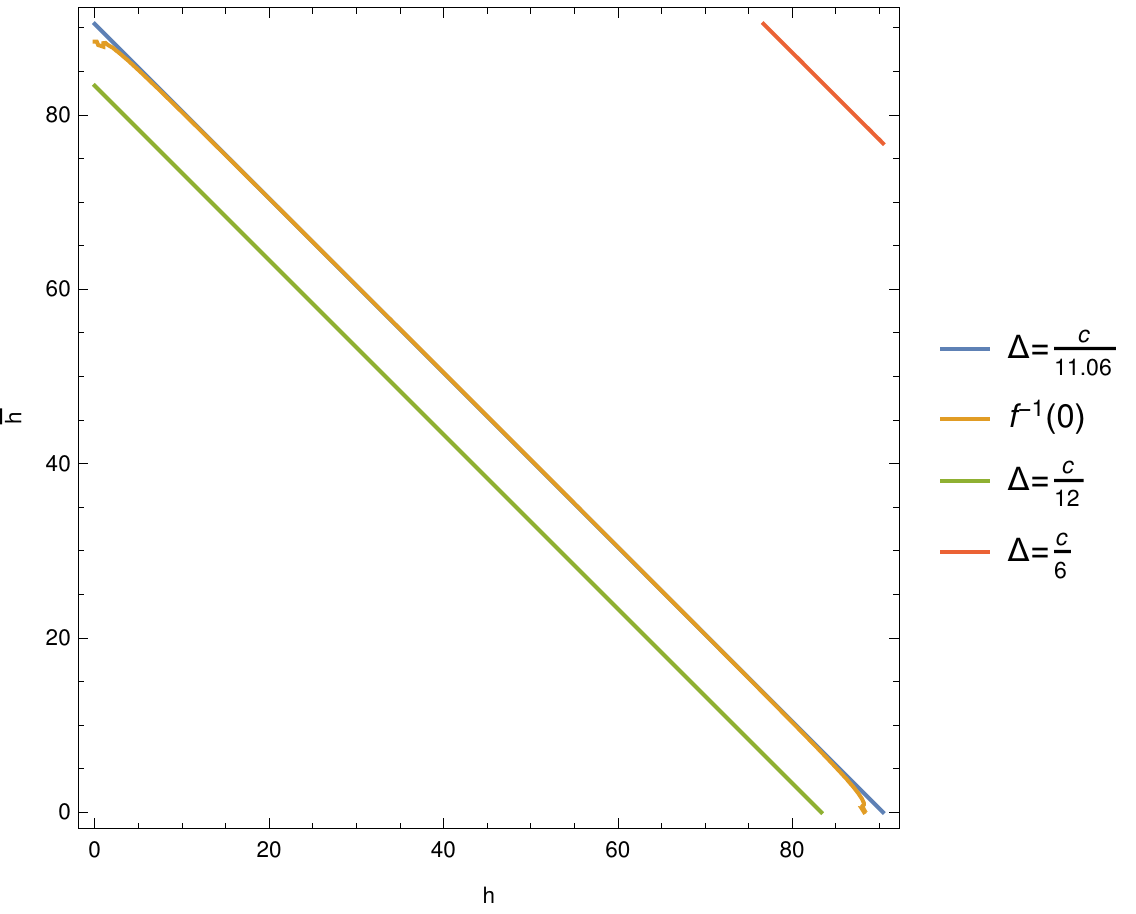}
\caption{The $(h,\bar{h})$ plane. The yellow line corresponds to the boundary that separates possible positive and negative contributions to the one-point function of an operator $\mathbb{O}$ with $h_\mathbb{O}= \frac{c}{12}$ in a theory with only positive diagonal OPE coefficients at $c = 2000$. There must be some contribution from the region bellow it. The region is contained in the region bound by $\Delta = h+\bar{h} = \frac{c}{10.38}$.  }
\label{fig:OnePtZeroLine}
\end{figure}
\noindent
From the analysis of the partition function we know that there must be a field with conformal dimension below the Hellerman bound. Since we can choose $\mathbb{O}$ in our present analysis, we can use the Hellerman bound to demand $h_\mathbb{O}\le \frac{c}{12}$ (forgetting about the possibility that the field obeying the bound may have spin!). Now, this actually can produce an improved upper bound by putting $h_\mathbb{O} = \frac{c}{12}$: We know that there has to be a field below $\mathfrak{f}^{-1}_{F_1}(0) \equiv \{(h^\star,\bar{h}^\star)|f_{F_1}(h^\star,\bar{h}^\star) = 0\}$ which cannot be the vacuum ($C_{0\Delta_i 0} = 0$). The numerical analysis then shows that $h^\star + \bar{h}^\star \leq \frac{c}{10.38}$ which shows that there has to be some field in the spectrum with $h_i+\bar{h}_i\lesssim \frac{c}{10.38}$. However, since we have a new bound we can perform the same procedure again and use $h_\O = \frac{c}{20.76}$ which then gives a slightly improved bound. We can repeat this procedure until it converge to some minimal upper bound. This bound is given by $\approx \frac{c}{11.05}$. Figure \ref{fig:OnePtZeroLine} shows the zero variety for $c=2000$. 

Note, that from \eqref{eq:1PtTorusBlock} onward we ignored all large $c$ corrections. Howevery, we shortly want to discuss their effect on the result. Firstly, they will generically shift the zero line by a sub-leading term (of order $\mathcal{O}(1))$, s.t. the bound $ c/11.05$ gets a sub-leading correction, too. Secondly, in principle it might happen that the sub-leading term gives rise to a second zero line. However, we expect that this does not happen. The number of zero lines is dictated by the polynomial power of $h,\bar{h}$ which is controlled by the numbers of derivatives in the functional and, hence, does not depend on $c$. Investigation of the torus blocks at $\mathcal{O}(1/c)$ order is presently under progress.\footnote{We want to thank Sridip Pal for his comments on this.}

Improving the above bound and constructing further bounds by exploring the space of possible functionals is left for future work.

\subsubsection{Constraints from the torus two-point function} \label{sec:T2ptfct}

The thermal two-point function can be written as 
\begin{align}
\left\langle \mathbb{O}(w,\bar{w}) \mathbb{O}(0,0) \right\rangle_\beta &= \text{\rm Tr}(\mathbb{O}(w,\bar{w}) \mathbb{O}(0,0))\\
& = \sum_{\Delta,s,\Delta',s'} \left|\bra{\Delta,S}\mathbb{O}\ket{\Delta',s'}\right|^2 e^{-\beta \left(\Delta-\frac{c}{12}\right)} e^{-2\pi\Im(w)(\Delta'-\Delta)+i 2\pi\Re(w) (s'-s)}\,,\nonumber
\end{align}
where $w = \Re(w) + i \Im(w)$ and  $\mathbb{O}(w,\bar{w}) = e^{\Im(w)\hat H +i\Re(w)\hat P}\,\mathbb{O}(0,0) \,e^{-\Im(w) \hat H - i \Re(w)\hat{P}}$, with $\hat{H} = 2\pi\left(L_0+\bar{L}_0 -\frac{c}{12}\right)$ and $\hat{P} = 2\pi\left(L_0-\bar{L}_0\right)$, and $s = h -\bar{h}$ is the conformal spin. 
Now, we want to assume that $\mathbb{O}$ is a scalar operator of conformal dimension $\Delta_\O<1/2$. 
In that case we can avoid UV divergences and do not have to worry about them.
Then we can use \eqref{eq:modInvNpointEasy} to construct the $S$-invariant function, i.e.
\begin{align}
g(\beta) :=& \int_0^{\frac\beta{2\pi}} d\Im(w) \int_0^{1} d\Re(w) \left(\frac{\beta}{2\pi}\right)^{\Delta_\mathbb{O}-1} \left\langle \mathbb{O}(w,\bar{w}) \mathbb{O}(0,0) \right\rangle_\beta\\
=& \left(\frac{\beta}{2\pi}\right)^{\Delta_\mathbb{O}} \bigg[\sum_{\Delta,s}  \left|\bra{\Delta,s}\mathbb{O}\ket{\Delta,s}\right|^2\, e^{-\beta \left(\Delta-\frac{c}{12}\right)}\label{eq:2ptSum}\\
&\qquad\qquad + \sum_{\Delta\neq\Delta',s} \left|\bra{\Delta,s}\mathbb{O}\ket{\Delta',s}\right|^2  e^{-\beta \left(\frac{\Delta+\Delta'}2-\frac{c}{12}\right)} \frac{\sinh\left(\beta\frac{\Delta-\Delta'}2\right)}{\beta\frac{\Delta-\Delta'}2}\, \bigg]\,.\nonumber
\end{align}

\noindent 
Note that the sum runs no longer over twice the full spectrum because the integral over $\Re(w)$ fixes $s=s'$. Now, applying $D_\beta^p$ we obtain the constraints
\begin{align}
0 = \sum_{\Delta,\Delta',S} \frac{\left|\bra{\Delta,S}\mathbb{O}\ket{\Delta',S}\right|^2 }{2\pi(\Delta-\Delta')} \sum_{n=0}^p \binom{p}{n} (\Delta_\mathbb{O}-1)^{p-n} \left(e^{-2\pi \Delta'}  B_n\left[\frac{\pi c}{6}-2\pi \Delta'\right]-(\Delta'\to\Delta)\right)\label{eq:2ptSum2}
\end{align}

\noindent
For $\Delta = \Delta'$ it is understood to take the limit $\Delta'\to \Delta$. The result is by construction symmetric under the exchange of $\Delta$ and $\Delta'$.\footnote{In \eqref{eq:2ptSum2} we have used: 
\begin{align*}
F_p\left[\left(\frac{\beta}{2\pi}\right)^x e^{\beta y}\right]&=\lim_{\beta\to2\pi} D_\beta^p \left(\left(\frac{\beta}{2\pi}\right)^x e^{\beta y}\right) \\
&= \sum_{n=0}^p \binom{p}{n} F_{p-n}\left[\left(\frac{\beta}{2\pi}\right)^x\right]F_{n}\left[e^{\beta y}\right]\\
&=  e^{2\pi y} \sum_{n=0}^p \binom{p}{n} x^{p-n} B_n[2\pi y]
\end{align*}}

Since $\left|\bra{\Delta,s}\mathbb{O}\ket{\Delta',s}\right|^2$ is always non-negative, whether some range contributes negatively or positively to the sum only depends on the functions
\begin{align}
h_p(\Delta,\Delta') :=  \frac{1}{2\pi(\Delta-\Delta')} \sum_{n=0}^p \binom{p}{n} (\Delta_\mathbb{O}-1)^{p-n} \left(e^{-2\pi \Delta'}  B_n\left[\frac{\pi c}{6}-2\pi \Delta'\right]-(\Delta'\to\Delta)\right)\,.
\end{align}

\begin{figure}\label{fig:ZeroLines}
\centering
\includegraphics[width=.48\textwidth]{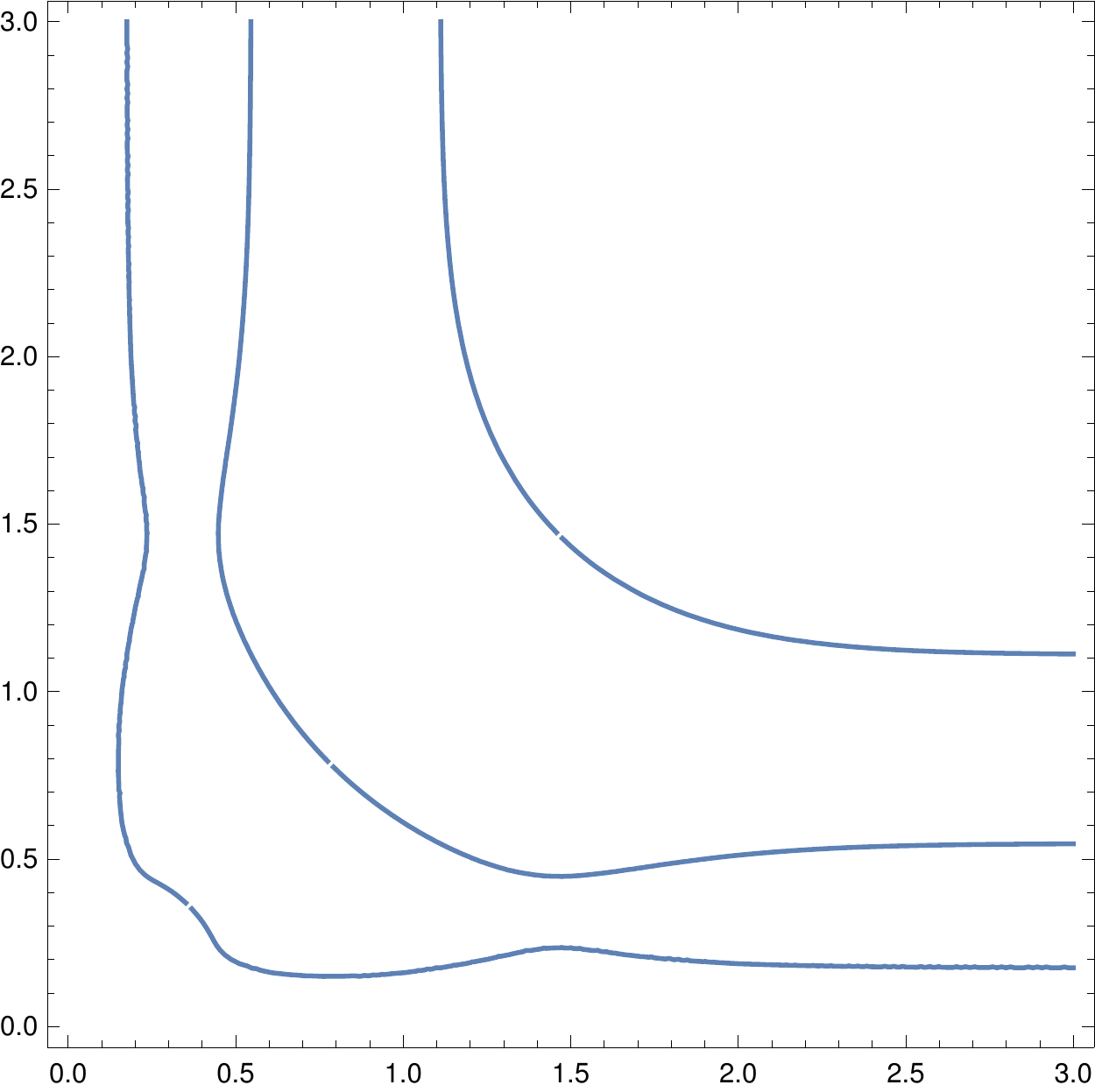}
\includegraphics[width=.48\textwidth]{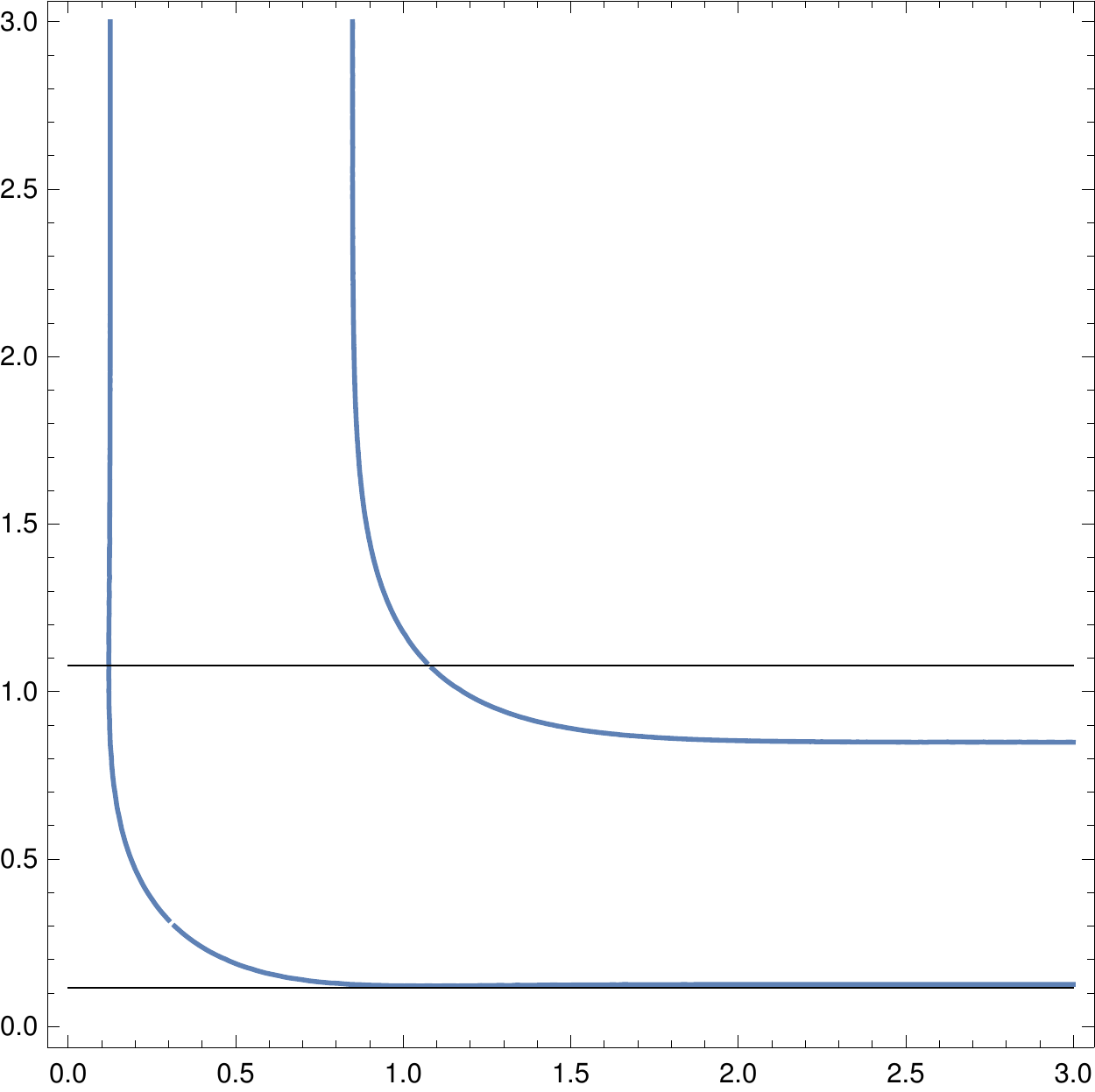}
\caption{Left: Zero lines of \eqref{eq:boundfrom3a1} in the positive quadrant for $c=3, \Delta_\mathbb{O} = 1/3,$ and $a_1 = 6, a_3=-3,a_5 =1$. Right: The bounds (y value of the black lines) (from \eqref{eq:boundfrom-a}) and zero lines in the positive quadrant for $c=3, \Delta_\mathbb{O} = 1/3,$ and $a$ as given in \eqref{eq:aFor2pt}. }
\end{figure}

\noindent 
One possibility to construct a bound follows the logic we saw in case of the partition function. 
The above sum over any linear combination 
\begin{equation}\label{eq:boundfrom3a1}
\sum_p a_p \, h_p(\Delta,\Delta')\,,
\end{equation} 
has to vanish. This means that regions in the positive quadrant of the $(\Delta,\Delta')$-plane that contribute positively have to cancel with the negative contribution from the complementary region. On the left of figure \ref{fig:ZeroLines} we show this regions, for $p\le 5$ and some specific values of $c$ and $\Delta_\O$, separated by zero lines extending from $(\infty,z_i)$ to $(z_i,\infty)$.

As an example we want to consider the linear combination  
\begin{equation}\label{eq:boundfrom-a}
h_3 (\Delta,\Delta') - a \, h_1(\Delta,\Delta')\,.
\end{equation} 

\noindent 
For $c>2.9425$ and 
\begin{align}\label{eq:aFor2pt}
 a = \frac{c^2 \pi^2}{36} + \Delta_\mathbb{O}^2 + \frac{c\pi}6 \left(3+2 \Delta_\mathbb{O} + \frac{6}{c\pi+6\Delta_\mathbb{O}}\right)
\end{align}
we obtain three non-compact regions separated by two zero lines as shown on the right of figure \ref{fig:ZeroLines}. 
For smaller values of $a$ a third zero line would intersect with the positive quadrant. 
For larger $a$ the upper zero line moves further to the upper right and, as we will see, would give weaker bounds. 
The bound on the central charge is chosen such that the lower zero line is not intersecting with the $\Delta = 0$ axis for any value of $\Delta_\O$.\footnote{For a specific value of $\Delta_\O$ the bound on $c$ can be improved and is at least numerically easy to compute.}
As a consequence only contributions from the region between the two other zero lines can compensate the remaining rest of the region. 
In other words, there have to be two fields $\xi_1,\xi_2$ with $(\Delta_{\xi_1},\Delta_{\xi_2})$ in the region bounded by the two zero lines and with $\bra{\xi_1}\O\ket{\xi_2} \neq 0$. 
This, in particular, gives the slightly weaker but more tangible constraint that there has to be at least one field $\xi$ with $\Delta_- < \Delta_\xi < \Delta_+$ that contributes to the sum. 
Here, $\Delta_-$ is given by the lowest $\Delta'$ value of the lower zero line, and $\Delta_+$ is given by the largest zero of $h_3(\Delta,\Delta) - a \, h_1(\Delta,\Delta)$. This is indicated by the two black lines on the right of figure \ref{fig:ZeroLines}. 
In particular, the value of $\Delta_+$ can be computed analytically and is given in Appendix \ref{app:2ptBounds}.
The bounds derived above are not expected to be optimal. 
The (numerical) search for an optimal functional is left for future work. 

\begin{figure}[t]\label{Figure:compact}
\centering
\includegraphics[width=.48\textwidth]{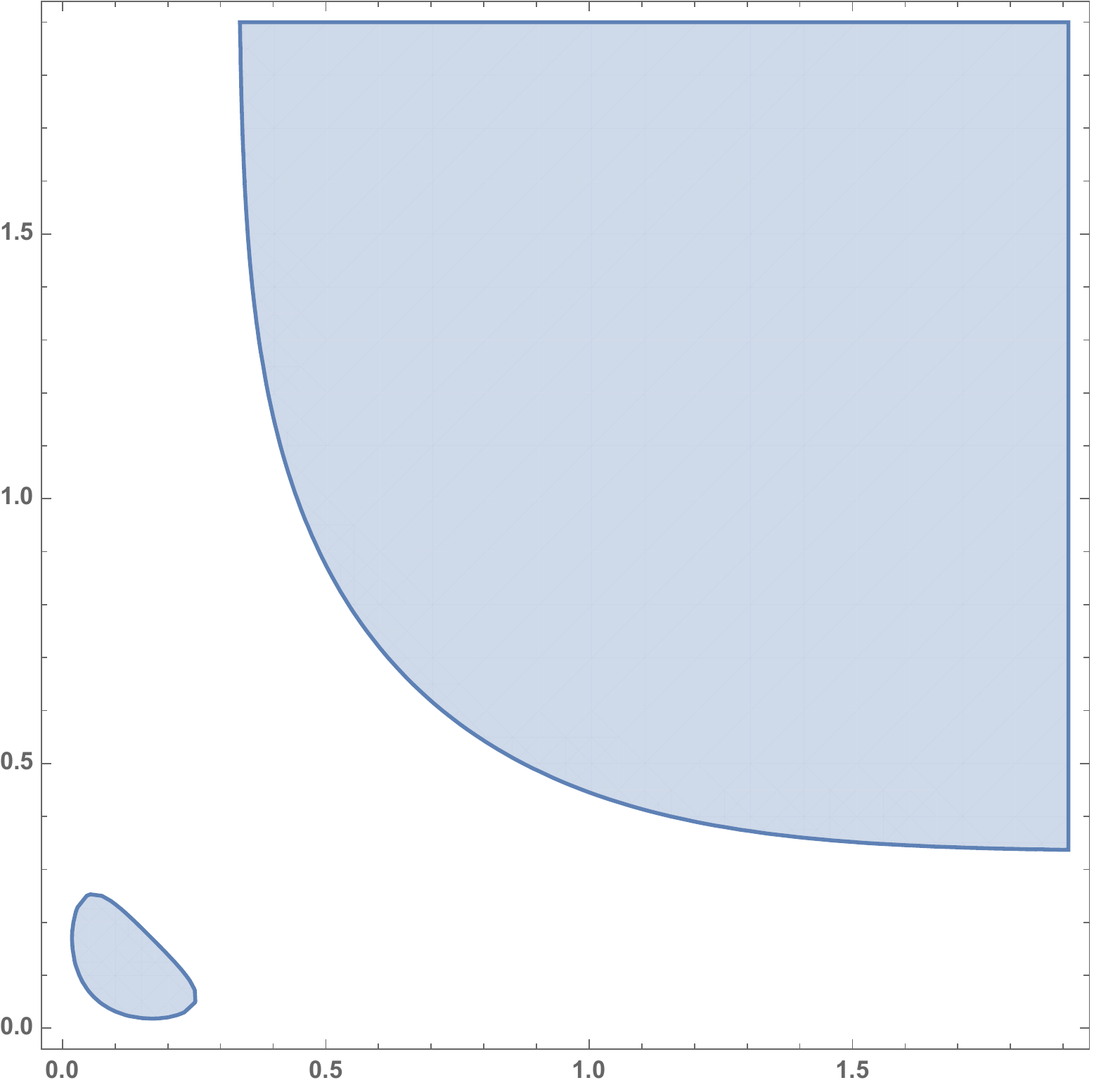}
\caption{ Plot of \eqref{eq:boundfrom-a} for $c=1, \Delta_\mathbb{O} = 1/3,$ and $a = 1$. The function is $<0$ within the shaded region. This exemplary plot, in particular, serves to show the appearance of an ''island``. However, here it appears together with a non-compact region, s.t. the proposed analysis cannot be applied. }
\end{figure}

However, we still want to discuss a particular important case that can appear. From the above discussion, it should be clear that the functional $F[\cdot]$ can be used to define a function 
\begin{align}
\mathfrak{f}_F:\qquad \mathbb{R}^2 &\to \mathbb{R}\\
 (\Delta,\Delta') &\mapsto \mathfrak{f}_F(\Delta,\Delta')\,,
\end{align}
by spreading it over the (generalized) sum \eqref{eq:2ptSum}. The regions given by $f_F^{-1}(0)$ then generically define boundaries between the regions with positive and negative contributions to the sum. In figure \ref{fig:ZeroLines} these are the drawn zero lines, where three (left) and two (right) of them have non-compact overlap with the positive quadrant $\mathbb{R}^2_{\ge0}$. If it is possible to construct a functional $F$ such that $\mathfrak{f}_F^{-1}(0)\cup \mathbb{R}^2_{\ge0}$ is compact and connected then this would define a finite ''island``, whose contribution has to compensate the contribution from its (non-compact) complement.\footnote{See an example of such an island but together with a non-compact region in figure \ref{Figure:compact}.} 

For a discrete spectrum, the contribution from the compact region is then finite and allows to obtain more concrete information about the matrix elements. This becomes most transparent if we assume that there exists a finite region in a theory with sparse enough spectrum such that only one particular $(\Delta_*,\Delta'_*)$ contributes. Then 
\begin{align}
(2-\delta_{\Delta_*,\Delta'_*})\left|\bra{\Delta_*}\mathbb{O}\ket{\Delta'_*}\right|^2 f_F(\Delta_*,\Delta'_*) &= - \sum_{\Delta,\Delta'\neq \Delta_*,\Delta'_*}\left|\bra{\Delta}\mathbb{O}\ket{\Delta'}\right|^2 f_F(\Delta,\Delta')\,,
\end{align}
where we dropped the dependence on conformal spin. Any explicit knowledge about some of the matrix elements on the r.h.s. can then be used to construct an inequality that constrains the matrix element $\left|\bra{\Delta_*}\mathbb{O}\ket{\Delta'_*}\right|^2$. So, assume we know the matrix elements for a specific set of states $\mathcal{S}$, then
\begin{align}
\left|\bra{\Delta_*}\mathbb{O}\ket{\Delta'_*}\right|^2 \ge - \sum_{\mathcal{S}}\left|\bra{\Delta}\mathbb{O}\ket{\Delta'}\right|^2 \frac{f_F(\Delta,\Delta')}{(2-\delta_{\Delta_*,\Delta'_*})f_F(\Delta_*,\Delta'_*)}\,.
\end{align}

\noindent 
The hardest part, here, is to construct the functional $F$, s.t. one obtains a compact zero variety in the positive quadrant.

\section{Conclusions}

In two dimensional CFTs the constraints coming from the S-modular transformation, while extremely informative, still require further exploration. While the extreme temperature bootstrap is mostly analytical, the intermediate temperature constraints usually require numerics. The constraints coming from the extreme temperature relations, give asymptotic behaviour of weighted spectral densities, and provide a perfect arena for the application of complex Tauberian theorems in order to obtain stricter bounds. This analysis, can further be refined using a Radamacher expansion, as carried out for the density of states in \cite{Loran:2010bd}. It will be interesting to write the Radamacher expansion for the moments of the OPE coefficients as a set of linear equations and bootstrap the CFT data by demanding  non-trivial solutions. Next, since the general transformation property of the \textit{n}-point torus correlator is known \eqref{eq:CorrTrafo}, one may generalize the one and two point function treatments (in \S\ref{Sec:onept-tor} and in \S\ref{Sec:twopt-tor} respectively) to higher point correlators. This will give rise to the asymptotic behaviour of higher moments of OPE coefficients, since now there will be more insertions of complete set of states in between the operators at different locations. This is going to be particularly interesting to understand if modular trafo allows for OPE moments to behave as required for obtaining chaotic behaviour (as characterized by the exponential growth of the unequal time commutator squared)  \cite{Foini:2018sdb, Murthy:2019fgs}. \\
Within the examples of the extreme temperature bootstrap, in \S\ref{Sec:pillow} and in \S\ref{Sec:zamu-rec}, we used the fact, that the crossing equation on the sphere, becomes S-transformation in a special conformal frame known as the pillow geometry. The recursive technology developed by Zamolodchikov \cite{Zamolodchikov1984, Zamolodchikov1987}, involves computation of coefficients of a q-expansion in the pillow geometry. These coefficients can be thought of as the density of states being exchanged in the s-channel of the conformal block, and depends sensitively on the external operator dimensions as well as the conformal dimension of the one exchanged. The time complexity of the Zamolodchikov recursion coefficients\footnote{Recently \cite{Ruggiero:2018hyl} have used the recursion coefficients to compute the Renyi entropy for two disjoint intervals.} calculated using the algorithm of \cite{Zamolodchikov1987} grow as $\mathcal{O}\left(n^3 ( \log n )^2 \right)$, when $n$ gets large\cite{kusuki3}. Hence it is advantageous to use the asymptotic bootstrap result for the large $n$. Though quite intricate the Virasoro kernel allowed us to get very detailed answer to the Zamolodchikov recursion density. We however,  focussed only on the leading part. It will be interesting to access the subleading contributions from the higher pole structures of the fusion kernel.  
The S-transformation constraints coming from the intermediate temperature analysis on the other hand yields bounds on the CFT data. The technology previously applied to the torus partition function can be generalized in a straightforward manner to torus correlators. For the two point function, \S\ref{sec:T2ptfct} the mathematical problem of optimization involves finding zero varieties that are compact and connected. While we have not solved the problem completely, it is clear that the technology of linear programming will be very useful. Additionally, we did not use the torus two point blocks in \S\ref{sec:T2ptfct} and used only the large $c$ blocks in \S\ref{sec:T1ptfct}, hence an immediate generalization is possible. Further, one may also extend the analysis to charged correlators, just as the methods reviewed in \S\ref{sec:Hellerman} for the partition function was generalized to the $\mathfrak{u}(1)$ case in \cite{Benjamin:2016fhe, Dyer_2018}, or to the cases with $\mathcal{W}_3$ algebra as done in \cite{Apolo_2017}. 

We also want to mention that there are more special points in the space of the modular paramter $\tau$ that could serve as candidates to develop new bootstrap techniques. In particular, the point $\tau = e^{2\pi i/3}$ is a fixed point under $S\cdot T$ and was e.g. analysed in \cite{Gliozzi:2019ewk}. There are also some conjectured transformation properties concerning the chemical potentials defining generalized equilibrium states with integrable charges \cite{Brehm:2019fyy}. They could also serve as starting points for bootstrap techniques and might lead to refined constraints on CFT spectra. 

\sectionlineB

\section*{Acknowledgements}
We would like to thank Sridip Pal and Corenelius Schmidt-Colinet for going through the draft. We will also like to thank Shouvik Datta, Axel Kleinschmidt, and Stefan Theisen for various illuminating discussions. D.D. will like to acknowledge the support provided by the Max Planck Partner Group grant MAXPLA/PHY/2018577, and the hospitality provided by AEI, Potsdam where part of this work was completed.

\appendix

\section{Fusion kernel for conformal blocks}\label{app:kernel}

In the parametrization $c = 1 + 6Q^2$, $Q = b+b^{-1}$, and $h = \alpha(q-\alpha)$, the relation between $s$- and $t$-channel confomal blocks is believed to be
\begin{equation}
F^{14}_{32}(h_t;1-z) = \int_C \frac{d\alpha_s}{2i} \,\mathbb{S}_{\alpha_s \alpha_t}\left[\begin{matrix}
\alpha_3 &\alpha_4\\ \alpha_1 & \alpha_2 
\end{matrix}\right] \cdot F_{12}^{34}(\alpha_s;z) \,,
\end{equation} 
with the fusion kernel
\begin{equation}\label{eq:Skernel}
    S_{\alpha_s\alpha_t}\left[\begin{matrix}
\alpha_3 &\alpha_4\\ \alpha_1 & \alpha_2 
\end{matrix}\right] = P(\alpha_i;\alpha_s,\alpha_t) P(\alpha_i;Q-\alpha_s,Q-\alpha_t) \int_{C'} \frac{ds}{i} \prod_{k=1}^4 \frac{S_b(s+U_k)}{S_b(s+V_k)}
\end{equation}

\noindent 
Here we use
\begin{align}
    &P(\alpha_i;\alpha_s,\alpha_t) \nonumber\\ =&\frac{\Gamma_b(\alpha_s+\alpha_3-\alpha_4)\Gamma_b(\alpha_s+Q-\alpha_3-\alpha_4)\Gamma_b(\alpha_s+\alpha_2-\alpha_1)\Gamma_b(\alpha_s+\alpha_1+\alpha_2-Q)}{\Gamma_b(\alpha_t+\alpha_1-\alpha_4)\Gamma_b(\alpha_t+Q-\alpha_1-\alpha_4)\Gamma_b(\alpha_t+\alpha_2-\alpha_3)\Gamma_b(\alpha_t+\alpha_2+\alpha_3-Q)}\cdot\\\cdot&\frac{\Gamma_b(2\alpha_t)}{\Gamma_b(2\alpha_s-Q)}\nonumber\,
\end{align}
and 
\begin{equation}
   S_b(x) = \frac{\Gamma_b(x)}{\Gamma_b(Q-x)} 
\end{equation}
with the Barnes double gamma function $\Gamma_b(x)$ satisfying $\Gamma_b = \Gamma_{b^{-1}}$ and the functional equation 
\begin{equation}
    \Gamma_b(x+b) = \frac{\sqrt{2\pi}b^{bx-\frac{1}{2}}}{\Gamma(bx)} \Gamma_b(x)
\end{equation}
and a similar equation for $b^{-1}$. For $\Re(x)\ge 0$ it admits the following integral representation
\begin{equation}
\log\Gamma_b(x) = \int_0^\infty \frac{dt}{t} \left[\frac{e^{-xt} - e^{-Qt/2}}{(1-e^{-bt})(1-e^{-t/b})} - \frac{\left(\frac{Q}{2}-x\right)^2}{2}e^{-t} - \frac{\frac{Q}{2}-x}{t}\right]\,.
\end{equation}

\noindent 
We also use the definitions
\begin{align}
    U_1 & = \alpha_1-\alpha_4   & V_1 & = Q-\alpha_s+\alpha_2-\alpha_4\\
    U_2 & = Q-\alpha_1-\alpha_4 & V_2 & = \alpha_s + \alpha_2 -\alpha_4\\
    U_3 & = \alpha_2+\alpha_3-Q & V_3 & = \alpha_t\\
    U_4 & = \alpha_2-\alpha_3   & V_4 & = Q-\alpha_t\,.
\end{align}

\noindent Finally, the integration contour $C'$ in the definition of $\mathbb{S}$ is from $-i\infty$ to $i\infty$ and passes through the tower of poles at $s^l_{i;m,n} = -U_i - mb -n/b$ and $s^r_{j;m,n} = Q-V_j+mb+n/b$, for $m,n\in \mathbb{Z}$. 

\subsection{Words on the spectrum}\label{app:anastruc}

The parametrisation used for the above definitions needs some specifications. Firs we want to specify $b$. For CFTs with $c>25$ one takes $b\in (0,1)$. In case of $c \in (1,25)$ one take $b$ to lie on the intercection of the unit circle with $\mathbb{C}^+ = \{z\in \mathbb{C}\,|\, \Re(z) > 0 \cup \Im(z)>0\}$. For $c>1$, any unitary theory only has positive conformal dimensions, $h\ge 0$. Then the spectrum, i.e. the range of conformal dimensions, splits into two pieces. The set of \textit{discrete} dimensions $h\in [0,\frac{c-1}{24})$, or $\alpha \in [0,\frac{Q}{2})$, and the set of \textit{continuous} dimensions $h \in [\frac{c-1}{24},\infty)$, or $\alpha \in \frac{Q}{2}+i\mathbb{R}$. The first set is dubbed discrete because the analytic structure of the kernel implies that only a discrete subset of dimensions in $h \in [0,\frac{c-1}{24})$ is actually supported in the $s$-channel. 

\subsection{Analytic structure of the kernel}

For generic $\alpha_t$ and $\alpha_i$ in the $t$-channel, the kernel has simple poles at
\begin{equation}
\begin{split}
    \alpha_{k,m,n} &= \gamma_k +mb +n/b\,, \quad \text{and } Q -\alpha_{k,n,m}\,, \quad \text{for } m,n\in \mathbb{N},\, k\in\{1,2,\dots 8\}\,,\\
    \gamma_1 &= \alpha_1 + \alpha_2, ~ \gamma_2 = \alpha_3+\alpha_4, ~(\text{six more obtained from $\gamma_{1,2}$ by }\alpha_i \to Q-\alpha_i)\,.
\end{split}
\end{equation}

\noindent 
In special cases, e.g. in case of pairwise identical external operators, the poles degenerate to \textit{double} poles. 

The pole structure, that basically follows from the choice of external fields, also determines the contour of integration in the $s$-channel. If all external fields are from the continuous spectrum $\alpha_i \in \frac{Q}{2} + i \mathbb{R}$ then it suffices to take $\alpha_s \in \frac{Q}{2}+i\mathbb{R}$, i.e. only continuum $s$-channel blocks contribute to the decomposition of the $t$-channel block. However, for external fields with $\alpha_1+\alpha_2<\frac{Q}{2}$ or $\alpha_3+\alpha_4<\frac{Q}{2}$, some of the pole cross the contour that gives the continuous spectrum. For the sake of analyticity in the parameters, these poles have to contribute to the decomposition by their residue. However, due to unitarity only $\alpha_s = \gamma_k + m b < \frac{Q}{2}$ can actually contribute. 

\section{Constraint function for the thermal one-point function} \label{app:OnePtConstraint}

The function $\mathfrak{f}_{F_1}$ to construct constraints from the thermal one-point function at large $c$ from the spinless operator $\mathbb{O}$ is given by
\begin{align} 
\mathfrak{f}_{F_1}(h,\bar{h},h_\mathbb{O},c) &=  \frac{e^{\frac{\pi}{6} (c+30 -12 (h+\bar{h}))}\Gamma(2 h)\Gamma(2\bar{h})}{3\left(1-e^{2\pi}\right)^3} \cdot \\
&\Bigg( - 6 e^\pi (h_\mathbb{O}-1) \pi \cdot \\
&\qquad \qquad\cdot\Big[ \,_2\tilde{F}_1\!\left(1-h_\mathbb{O},h_\mathbb{O},2h,\tfrac{1-\coth(\pi)}{2}\right)\,_2\tilde{F}_1\!\left(2-h_\mathbb{O},h_\mathbb{O},2\bar{h},\tfrac{1-\coth(\pi)}{2}\right) + \nonumber\\
& \qquad \qquad +\,_2\tilde{F}_1\!\left(2-h_\mathbb{O},h_\mathbb{O},2h,\tfrac{1-\coth(\pi)}{2}\right)\,_2\tilde{F}_1\!\left(1-h_\mathbb{O},h_\mathbb{O},2\bar{h},\tfrac{1-\coth(\pi)}{2}\right)\Big]\nonumber \\
&\quad+ \Big[12 h_\mathbb{O}\pi \cosh(\pi) + \left\{-6h_\mathbb{O} -c\pi + 12 \pi (h_\mathbb{O} + h + \bar{h}-2)\right\} \sinh(\pi)\Big] \,\cdot \nonumber\\
&\quad\,\cdot\, \,_2\tilde{F}_1\!\left(1-h_\mathbb{O},h_\mathbb{O},2h,\tfrac{1-\coth(\pi)}{2}\right)\,_2\tilde{F}_1\!\left(1-h_\mathbb{O},h_\mathbb{O},2\bar{h},\tfrac{1-\coth(\pi)}{2}\right) \Bigg)\,,\nonumber
\end{align}
where $_2\tilde{F}_1$ is the regularized hypergeometric function 
\begin{equation}
    _2\tilde{F}_1(a,b,c,z) := \frac{_2F_1(a,b;c;z)}{\Gamma(c)}
\end{equation}

\section{Bounds from the two-point function on the torus}\label{app:2ptBounds}

The upper bound (u.b.) is the largest zero of \eqref{eq:boundfrom3a1} with $\Delta = \Delta'$ and $a$ given as in \eqref{eq:aFor2pt}. It is given by 
\begin{equation}
\Delta^{u.b.} = \frac{18+3c\pi+18\Delta_\mathbb{O} + \sqrt{c^2\pi^2 + 36 \left(1+\Delta_\mathbb{O}\right)\left(5+\Delta_\mathbb{O}\right) + 12 c\pi \left(3+\Delta_\mathbb{O} + \frac{12}{ c\pi + 6\Delta_\mathbb{O} }\right)}}{24\pi}\,. 
\end{equation}

\noindent 
Note that for $c>\frac{3(\sqrt{11}-3)}{\pi} \approx 0.3$ this bound is larger than the bound that has been computed from the invariance of the partition function for all positive values of $\Delta_\mathbb{O}$. As such it does not give any stronger restriction on the general spectrum of the theory.

\bibliography{lit}

\end{document}